\documentclass[preprint,12pt,sort&compress]{elsarticle}
\usepackage{amsmath}
\usepackage{amssymb}
\usepackage[titletoc,toc,title]{appendix}
\usepackage{graphicx}
\usepackage{float}
\usepackage{bm}
\usepackage{xcolor}
\usepackage{booktabs}
\usepackage{hyperref}
\usepackage{diagbox}
\usepackage{braket}
\usepackage{nicefrac}
\usepackage{dsfont}
\usepackage{comment}

\renewcommand{\vec}[1]{\textrm{\textbf{#1}}}
\def\ie {{\it i.e.} }
\definecolor{river}{HTML}{517C96}

\journal{Computer Physics Communications}

\begin{document}

\begin{frontmatter}

\title{Dynamic Simulations of Strongly Coupled Spin Ensembles   for Inferring  Nature of Electronic Correlations from Nuclear Magnetic Resonance}

\author[a]{Charles Snider}
\author[a,b]{Stephen Carr}
\author[a,b]{D. E. Feldman}
\author[c]{Chandrasekhar Ramanathan}
\author[a,b]{V. F. Mitrovi\'c}

\address[a]{Department of Physics, Brown University, Providence, Rhode Island 02912-1843, USA}
\address[b]{Brown Center for Theoretical Physics and Innovation, Brown University, Providence, Rhode Island 02912-1843, USA.}
\address[c]{Department of Physics and Astronomy, Dartmouth College, Hanover, NH 03755, USA}

\date{\today}

\begin{abstract}

We develop an efficient package for the simulation of nuclear magnetic resonance spin echo experiments to study the effects of strong electronic spin correlations on the dynamics of the nuclear spin ensemble. A mean-field model is used to study correlated electronic phases through their hyperfine interaction with nuclear spins. We explore the dynamics of the interacting nuclear ensemble and discuss the key behaviors of the system.  In particular, we classify the types of temporal asymmetry that the interaction induces in the system as well as a pulse-dependent shift in the spectral domain.  Using these results, we discuss how careful measurement of the pulse-dependent shift can be used to extract information about the anisotropy of the electronic interaction and how these results represent a novel tool for the examination of exotic NMR signatures in strongly correlated materials.  Finally, we review specific aspects of the simulation package developed for our exploration and give explicit examples where package can be used to infer range and anisotropy of electronic correlations.  In particular, we discuss its structure, accuracy, and the technical merits of the various approximations used to model the nuclear spin ensemble.

\end{abstract}


\end{frontmatter}


\clearpage
\noindent{\bf PROGRAM SUMMARY}

\begin{small}
\noindent
{\em Program Title: Spin Echo Sim}                                          \\
{\em CPC Library link to program files:} (to be added by Technical Editor) \\
{\em Repository link:}  \url{https://github.com/charlessnider/Spin_Echo_Sim} \\
{\em Code Ocean capsule:} (to be added by Technical Editor)\\
{\em Licensing provisions:} GPLv3 \\
{\em Programming language:} Julia, C++, CUDA             \\                     
{\em Nature of problem:} 
Nuclear magnetic resonance studies of strongly correlated electronic materials can result in unexpected or anomalous echo shapes \cite{Mitrovic2008}.
Such shapes can be explained by a long-range nuclear spin-spin interaction~\cite{carr2022, Rao23}.
However, standard NMR simulations packages are not designed to simulate such large, interacting spin systems.
A simulation package designed around the efficient dynamic simulation of large spin ensembles ($n > 100,000$) whose interactions can be represented in the mean field limit therefore represents an unexplored yet potent application of numerics to the study of strongly correlated electrons systems in the context of NMR. \\
{\em Solution method:}  
 Our open-source CUDA code provides a highly efficient package for the simulation of NMR experiments in the presence of strongly correlated electrons represented in the mean field limit. 
 With full GPU implementation, the simulation performance is hundreds of times faster than standard CPU or hybrid CPU/GPU techniques.
 The package can accommodate any single-spin or mean field Hamiltonian.
 A more accessible and easily-adjustable (but less efficient) Julia version of the simulation is included in the code base, as well as tools for the automatic creation of parameter input files for the CUDA simulation. \\

\end{small}

\section{INTRODUCTION}
\label{sec: intro}

Strongly correlated electron systems have been a central pillar of research in condensed matter physics in recent decades.
A great variety of unique and complex physical phenomena, challenging both to theoretically model and to experimentally identify, emerge in the presence of strong electronic correlations.
Much experimental effort has gone into charting out the location of and conditions for novel electronic and magnetic phases arising from these strong correlations, and theoretical explorations have brought to the forefront the potential for a wide variety of new phenomena.
For example, triplet superconductivity  \cite{RevModPhys.75.657,RevModPhys.77.1321, Alloul_2015, YangTrip21,Benhabib:2021aa,Sim:2022aa,PhysRevB.107.014506} and Fulde-Ferell-Larkin-Ovchinnikov (FFLO) superconductivity \cite{PhysRev.135.A550, Matsuda_2007,Mitrovic08,Mitrovic2010,Mayaffre14} represent some of the most exotic phases, and they are still not well understood from experimental measurement.
The fractional quantum Hall effect also remains a major source of enigmatic strongly correlated phases \cite{RevModPhys.71.S298,RevModPhys.75.1101,Jain2015,feldman:2021}.
Other magnetic systems offer multiple strongly correlated states ranging from putative spin liquids \cite{Savary_2017, RevModPhys.89.025003, RuCl3} to analogues of superfluids \cite{FK1990,FK1991,RevModPhys.94.035004,RevModPhys.71.S318,PhysRevB.107.085403,PhysRevX.13.021037}.

Nuclear Magnetic Resonance (NMR) is an important tool in the study of strongly correlated phases because of its unique access to electrons via their relatively weak hyperfine interaction with nuclear spins \cite{Rigamonti98,Alloul,BERTHIER2017331,PhysRevB.76.115335,PhysRevB.78.014504,Vachon2011,PhysRevResearch.5.L012017}. Because the characteristic frequencies of the nuclei are much smaller than those of electrons, NMR offers a probe of the electronic states which does not perturb the state itself. 
One of the most powerful static measurements for the analysis of the electronic spin states in NMR is the Knight shift.  The Knight shift refers to an overall shift in the nuclear frequency spectrum caused by the magnetization of conduction electrons and is driven by the hyperfine interactions between the electrons and nuclei \cite{PhysRev.76.1259.2,AbragamBook}. The manner in which the Knight shift changes relative to temperature through a superconducting transition can yield valuable information about the electronic spin susceptibility \cite{Mitrovic08}.  For example, observation of a lingering Knight shift below $T_c$ yields information about the type of pairing present in a   superconductor \cite{PhysRevB.56.R505,Alloul_2015}. Such information must be carefully extracted however, because heating effects \cite{Ishida_2020, Pustogow2019} can easily be misinterpreted as a residual Knight shift~\cite{MUKUDA19999, PhysRevB.56.R505, ISHIDA2008}.
Additionally,  formation of the FFLO state relies on the polarizability of the superconducting electrons and consequently a change in the spin susceptibility and Knight shift \cite{Matsuda_2007, Mitrovic2008, PhysRevLett.97.117002, Mayaffre14}. However, these  superconducting electrons  are likewise sensitive to pulse-induced heating \cite{Mitrovic2008}.  

Apart from issues with sample heating, measurements of these strongly correlated phases bring a variety of other difficulties  with them. The emergence of unusual spectral features in NMR free induction decay (FID) or spin echo experiments is not uncommon \cite{Mitrovic2010, PhysRevMaterials.7.084409}. Asymmetric and multi-peak behavior can occur in the time domain, an oddity for NMR signals.  Such time domain asymmetry is typically attributed to experimental failure.  We propose that both of these behaviors -- the spectral distortions (asymmetric peaks) and spectral position (Knight shifts) -- may instead be indicative of a strongly correlated electronic mode inducing unique dynamics in the nuclear spin ensemble. Testing this proposal requires development of efficient numerical simulation of strongly interacting many spin systems. Here, we  describe such simulation package and benchmark its performance. We point out that this is a distinct simulation package that computes NMR observables in strongly interacting spin systems. Specifically, we have employed this package to simulate spin-echo shapes of  interacting nuclear spins in material with strong electron spin correlations.  In such materials,  the hyperfine interaction conveys long-range coupling between nuclear spins and dominates over all other nuclear spin-spin couplings \cite{Lu17, Vachon08}.  

This   unique package designed to directly investigate strongly correlated phases of matter,  
was  successfully employed   to  probe the electronic susceptibility via the variation of the pulse strength and applied field orientation, which has direct applications in sensing and characterizing emergent electronic phases ~\cite{carr2022,Rao23, Mitrovic2010,Mitrovic08, Mitrovic2008}. In particular, we demonstrated that from a simple spin-echo NMR experiment, one can deduce the spatial extent of electronic correlations and the total interaction  strength ~\cite{Rao23}, as well as the anisotropy of electronic correlations ~\cite{carr2022}.
  Simulations of the dynamics of weakly interacting and/or single nuclear spin systems, are less time consuming. An example of an up-to-date, open-source software that can simulate the time evolution of weakly interacting nuclear spins, is PULSEE \cite{Candoli23}. It is based on the quantum mechanical description of magnetic resonance and can simulate the time evolution of weakly nuclear spins in a wide variety of configurations encountered experimentally. Although, PULSEE was designed to handle non-interacting, single-spin systems in solids dominated by the Zeeman and quadrupolar interactions, the software can handle relevant coupling with other nuclei and/or electrons, simulating the evolution of single-spin systems subject to different pulse sequences. Dynamics of  a lattice of nuclear spins in a strong Zeeman field interacting via secular dipolar Hamiltonian was investigated in detail in \cite{Cho05}. Even though it was found that dipolar  interaction and arrangement of the spins on the lattice  plays a significant role in the development of higher-order correlations, dipolar interactions are often negligible in strongly correlated electron materials. 

To recap, there are essentially three assumptions in our work that made existing spin simulation tools and software inadequate to handled all important requirements of the problem. Some existing packages are designed for specifically the  ${1 \over {r^3}}$ dipole-dipole  interaction and rely on an Ewald technique for a convergent summation for periodic spin lattices. As we wanted to study longer range interactions, this was not appropriate.  In other cases, the software was designed for explicit propagation of the entire Hilbert space (e.g. SIMPSON \cite{Bak2000}). 
 Due to the exponential growth in the Hilbert space with the spin
count, usually only systems with up to 20 spins (for compute limits) or 30 spins (for
memory limits) can be handled, which is not large enough for us to study long-range interactions in 2-dimensions (2D) without artificial periodic reflections.
Finally, we require explicit time-propagation and application of a time-dependent Hamiltonian for the spin echo. Some codes are designed for computing ground states of spin
lattices only via e.g. DMRG or quantum Monte Carlo.
There are certain examples of papers on mean-field treatments of large 2D and 3D spin-
networks, but they are all related to Ising
spin models, used mostly for problems of spin ices.
Therefore,  to directly investigate strongly correlated phases of matter as was done in \mbox{Refs. \cite{carr2022,Rao23}}, we had to develop   simulation package of many-interacting spins, such as that described in this manuscript.

We replicate the dynamics of the nuclear ensemble in the presence of electron-mediated long-range spin-spin interactions. We demonstrate how such interactions can result in the exact pulse-dependent anomalous Knight shifts which have been measured in controlled environments  with no  potential for heating induced effects  \cite{Vinograd2021}.  Furthermore, through a careful analysis of the Knight shift, measured under different effective pulsing conditions, we outline how information about the anisotropy of the electronic spin susceptibility and   relative length scale of the electronic correlations can be extracted from the anomalous shifts.
We have previously demonstrated how the use of even simple models in tandem with numerical simulations represents a new methodology by which strongly correlated materials can be probed and understood ~\cite{carr2022,Rao23}. 
This work serves as a companion to this previous publication, and focuses on providing a much more thorough description of the computational implementation.

In Section~\ref{sec:model} we introduce the model for mean-field spin-spin interactions and time propagation of a large ensemble of nuclear spins. In Section~\ref{sec:results} we present results from this model for various choices of spin-spin interaction strength, range and anisotropy, which are inherited from the underlying electronic phase. Furthermore, we discuss how our findings compare to explicit experimental findings in unconventional superconductors. In Section~\ref{sec:simulation} we explain details of our numerical implementation, particularly our use of CUDA GPU acceleration,  and examine convergence of our model with respect to all simulation parameters. In Section~\ref{sec:conclusion} we summarize our key conclusions and discuss future directions.

\section{MODEL} 
\label{sec:model}

Here, we will investigate a lattice of nuclear spins interacting via a long range coupling (more than a few lattice lengths). We point out that the influence of highly local inter-nuclear interactions, such as through-bond $J$ couplings (pseudo-exchange)  or magnetic dipolar couplings, are not our focus as this is address in \mbox{Ref. \cite{Candoli23}}.  The effects of short-range interactions are well studied and well understood in NMR, whereas our exploration will be into the effect and detection of long range inter-nuclear interactions produced by electronic correlations.  To that end, to isolate the effect of long rage interactions between the spins, we do not include these conventional terms in the Hamiltonians studied here. 

To reduce computational complexity and efficiently model a large ensemble of spins, we employ a mean field approximation  for the electron mediated  interaction between nuclear spins. Using a more exact approach becomes computationally intractable at the length scales we wish to study. As our model focuses on long-range couplings, spatial averaging of the interaction is expected to occur and improve the mean field approximation.  We will still consider the influence of spatial variations by considering a smoothly varying local field.

\subsection{Mean field Hamiltonian}
A generic model for a system of interacting nuclear and electronic spins is
\begin{equation}
    H = \sum_{i} H_N(\vec{I}^{(i)}) 
    + \sum_{k,l} H_E(\vec{S}^{(k)},\vec{S}^{(l)})
    + \sum_{i,k} H_{hf}(\vec{I}^{(i)},\vec{S}^{(k)})
\end{equation}
where $\vec{I}^{(i)}$ is the nuclear spin operator for atomic site $i$, $\vec{S}^{(k)}$ is the electron spin operator for electron label $k$, $H_N$ is a generic nuclear Hamiltonian, $H_E$ a generic   electronic spin Hamiltonian (with parameters usually notated by $J_{l,k}$) and $H_{hf}$ the hyperfine interaction between nuclear and electronic spins.
Dealing with such a model exactly is of course intractable at large system sizes, as for $N$ atoms of spin $S$ and $M$ electrons would require a basis of size $(2S+1)^N 2^M$.

To gain a tractable model, we integrate out the electronic degrees of freedom ($\vec{S}^{(i)}$), leaving behind an effective interaction between nuclear spins:
\begin{equation}
     H' = \sum_{i} H_N(\vec{I}^{(i)}) 
    + \sum_{i,j} J_{NN}(i,j) (\vec{I}^{(i)},\vec{I}^{(j)}) + H_{D}.
\end{equation}
The term $H_{D}$ captures all   dissipative interactions affecting  the nuclear degrees of freedom after integrating out the 
 electrons, such as  dissipation of energy to the external bath of electrons.  
For now, we will ignore these terms, but we describe their treatment via a Linbald form of the Markovian master equation in Sec.~\ref{sec:liouville}.

Although an explicit hyperfine interaction no longer exists in the simplified Hamiltonian, the effective nuclear-nuclear interaction $J_{NN}$ is mediated by the hyperfine interactions between nuclei and electrons.
To fully capture the effect of the removed electrons, the interaction should couple nuclear spins across time and space, but we will assume that the important physics is captured by the instantaneous response. 
This assumption is justified by the fact that any electron interaction occurs on the times scale thousand time faster then that of the nuclei. Therefore, any electron mediated interaction can be treated as instantaneous in the nuclear spin reference frame to yield: 
\begin{equation}
J_{NN}(r_i - r_j, t_i - t_j) \approx J_{NN}(r_i - r_j) \delta_{t_i,t_j} .   
\end{equation}

We can reduce the complexity of the model even further by making a mean-field approximation.
The sum over pairs of nuclear spins is then replaced by a single onsite Hamiltonian for each nuclear spin $i$:
\begin{align}
    H^{(i)} = -\gamma \vec{I}^{(i)} \cdot \vec{M}^{(i)}
\end{align}
where \textbf{M}$^{(i)}$ is the effective magnetic (mean) field felt at site $i$ due to other nearby nuclear spins.
The magnitude of $M$ depends on both the strength of the coupling between the electrons and the nuclei and the distance between the nuclei.
We write the effective field \textbf{M}$^{(i)}$ as a sum over couplings to nearby nuclei, $j$,
\begin{align}
    \label{eq:eff_field}
    M^{(i)}_d = \alpha_d \sum_{j \neq i} f(|\vec{r}_i - \vec{r}_j|)\langle I^{(j)}_d \rangle \quad \textrm{for $d = x, y, z$, }
\end{align}
where $f$ denotes  the localization function, that captures the spatial extent of the nuclear interaction, $\alpha_d$ represents the axis-dependent interaction parameter, and average in  $\langle I^{(j)}_d \rangle$  is over entire nuclear ensemble. 
Here 
we have made a further assumption that the effective nuclear-nuclear interaction $J_{NN}$ has no cross terms (e.g. $I_x I_y$), otherwise we would have to use a tensor form for $\vec{M}$ to allow for them.
We point out that $\vec{M}^{(i)}$ facilitates a back-action of the nuclei on themselves; where the magnitude of the action on a nuclear spin $i$ depends on the states of nearby electronic spins.  The full Hamiltonian for a given spin $i$, after including the Zeeman term, becomes
\begin{align}
    \label{eq:hamiltonian}
    H^{(i)} = - \nu_i I_z^{(i)} - \gamma \sum_{d = x, y, z} \alpha_i I^{(i)}_d M^{(i)}_d
\end{align}
with $M_d^{(i)}$, as defined in Equation \ref{eq:eff_field},  that captures the main features of the electronic spin susceptibility in materials and thus varies from one correlated electronic phase to the other.  Importantly, the spatial range of  $M_d^{(i)}$  is not determined by the distance between the nuclear spins (as is
the case in dipole-dipole term) but by nature of electronic susceptibility itself.   
We point out that this mean field approach is the most appropriate to capture effect electron mediated interactions on the nuclear spin dynamics. Other approaches such as the dynamical mean field  (DMF) theories were only successfully applied when dealing with dipole-dipole system \cite{PhysRevB.72.054427,PhysRevResearch.5.043191}. Moreover, cluster expansion was useful for the central spin model coupled to the spin bath - random bath \cite{https://doi.org/10.1002/adts.202100254, PhysRevB.75.125314,PhysRevB.102.245303,Jing:2018aa,PhysRevB.109.224305,PhysRevLett.132.250401}.  

 Nevertheless, the  limitations of the mean field approach adopted here  are well understood  and do not compromise the regime of validity of the nuclear spin dynamics dictated by the electron mediated interactions. First, the electronic degrees of freedom that mediate the nuclear-nuclear interaction evolve on timescales orders of magnitude faster than those of the nuclei, making an instantaneous description of the effective interaction appropriate. Furthermore, the finite interaction range encoded in the localization function $f(r,\xi)$ ensures that each nuclear spin couples to many neighbors, so that local fluctuations self-average and the dominant effect of the electronic environment is a coherent, anisotropic feedback field acting on individual nuclei. In this parameter regime, the mean-field Hamiltonian captures the leading-order nuclear spin dynamics while avoiding the prohibitive complexity of fully correlated treatments, which are neither necessary nor practical for the questions addressed here.

Our model has several tunable parameters that dictate the behavior of the system. The axis-dependent interaction parameter $\alpha_d$ represents both the strength of the hyperfine interaction and the susceptibility of the electronic structure (that is to say, the magnetic anisotropy  of the electronic structure).  
The localization function $f$ captures the spatial extent of the nuclear interaction, and is dependent only on the range of the electrons' magnetic susceptibility.  The combination of the two defines the overall ``strength'' of the interaction.

The localization function $f$ can be set to have any structure necessary to mimic the systems being studied.  For this work, we will use a Gaussian form defined as
\begin{align}
    f(|\vec{r}_i - \vec{r}_j|) \equiv f(r,\xi) = e^{-\left( \nicefrac{r}{\xi} \right)^2}.
    \label{eq:f_gaussian}
\end{align}
The Gaussian localization aims to mimic an electronic excitation with a gap; other potential localization schemes, such as a power law $f(r,p) = \nicefrac{1}{r^p}$ or an RKKY type $f(r,\xi) = x^4(x\cos x - \sin x)$ for $x = \nicefrac{2\xi}{r}$ could be used to model systems such as one without a gap  or   simple metal,  respectively \cite{theory_paper,PhysRev.106.893, PhysRev.96.99}.
We point out that the  localization function represent a spatial extent of nuclear-nuclear interactions that is mediated by the electrons. Therefore, it depicts the properties of the open-quantum system (nuclear spins coupled to the electronic bath). Gapped electronic excitations, as considered in\mbox{Eq. \ref{eq:f_gaussian}}, do not affect the memory of the nuclear spin system. Therefore, the Gaussian form correctly depicts open-quantum with memory; while, exponential functions would have Markovian no-memory description.

\subsection{The Nuclear Ensemble}
\label{sec:methods_ensemble}
We consider only a single material system for this study, a two dimensional square lattice of spin-$\nicefrac{1}{2}$ nuclei with periodic boundary conditions.
Restricting ourselves to two dimensions allows for longer length-scales of the lattice without excessive computation time, permitting the study of both short-range and long-range interactions.
We generally ignore temperature dependent effects and initialize this system in a ground state with all spins pointing along the $z$ axis.
To account for temperature effects more accurately, a more sophisticated approach like those used in open quantum systems could be used to define our starting state \cite{RevModPhys.88.021002}.

Due to small variations in the local nuclear environment, nuclear spins will have a distribution of precession rates under an applied magnetic field.
We use a Lorentzian distribution for the spectrum with a linewidth of $\Gamma$, centered at a resonant frequency of $10$~MHz.
However, as we work in the rotating frame this central frequency will be shifted to $0$~MHz and generally ignored.
We do not need to pick a specific linewidth $\Gamma$, because the Hamiltonian (Eq.~\ref{eq:hamiltonian}) has a scaling symmetry.
That is to say, multiplying every term in the Hamiltonian by a fixed amount does not change the spin dynamics, only the rate at which they occur is affected.
Therefore, we report all frequencies in units of $\Gamma$ and all times in units of $\Gamma^{-1}$.
To assign a frequency to each spin, we sample within the range $[-5\,\Gamma, 5 \,\Gamma]$ from a Lorentzian (Cauchy) distribution.

\subsection{Pulses and Time Evolution}
We assume instantaneous RF pulses, which allowing us to ignore the interaction and Zeeman precession during the pulse.  Under this assumption, we can  implement the pulses as simple rotation matrices on each spin's density matrix
\begin{align}
    \rho_f = e^{i R \varphi/\hbar} \rho_0 \, e^{-i R \varphi/\hbar},
\end{align}
where $R$ is the appropriate spin matrix for the axis of rotation (typically $I_x$ or $I_y$), $\varphi$ the flip angle, $\rho_0$ the density matrix before the pulse, and $\rho_f$ its density matrix after.

The time evolution of the density matrix is dictated by
\begin{align}
    \frac{d \rho}{dt} = \frac{i}{\hbar}[\rho, H].
\end{align}
We can define the density matrix 
 reference frame as
\begin{align}
    \rho_R = U^\dag \rho\, U \longrightarrow \rho = U \rho_R \, U^\dag \, , 
\end{align}
with $U = \exp(-i\nu I_z t/\hbar)$.  Making this replacement, we find
\begin{align}
    \frac{d\rho_R}{dt} = \frac{i}{\hbar}[\rho_R, U^\dag H \, U + \nu I_z] \, .
\end{align}
We recover an ``effective'' Hamiltonian in the rotating frame, $H_R$ such that
\begin{align}
    \frac{d\rho_R}{dt} = \frac{i}{\hbar}[\rho_R, H_R] \quad \textrm{with} \quad H_R = U^\dag H \, U + \nu I_z \, .
\end{align}
Using this transformation, our Hamiltonian from Equation \ref{eq:hamiltonian} becomes
\begin{align}
    \label{eq:rot_ref1}
    H^{(i)}_R = -\left( \nu_i - \nu \right)I_z^{(i)} - \gamma \sum_{d = x,y,z} \alpha_d U^\dag I_d^{(i)} U M^{(i)}_d \, .
\end{align}
The density matrix $\rho_R$ still undergoes Larmor precession, but at a frequency $\Delta \nu \equiv \nu_i - \nu$, which is typically on the order of kHz rather than MHz, resulting in a period on the same order as the experiment time ($2 \tau$). 
By eliminating the high-frequency Larmor precession, we drastically reduce the temporal resolution required in our calculation.  

We model the simple spin echo experiment by applying a $\pi/2$ and a $\pi$ pulse separated by an echo time $\tau$.
The first pulse moves the $z$-polarized spins into plane at time $t=0$, while the second pulse at $t=\tau$ attempts to invert their accumulated phase in order to generate an ``echo'' at $t=2\tau$.
Although we will refer to these two pulses as $\pi/2$ and $\pi$ pulses, sometimes we will consider the effect of spin rotation angles that differ from these values, indicated by a simulation parameter $\theta$.
We apply both pulses along the $x$ axis of the rotating frame, resulting in a net magnetization with only a $y$-component in the absence of spin-spin interactions.
Although the choice of ($x,x$) for the pulse axes means the $y$-component of the magnetization will be negative at the echo, we report only the absolute value of the $y$-magnetization, $|M|$.

\subsection{Interactions in the Rotating Frame}
\label{sec:rot_frame}

The transition to the rotating frame may induce high-frequency behavior in the interaction term, as the planar interaction terms $\alpha_x$ and $\alpha_y$ are constant in the laboratory frame but not the rotating frame. 
As we shall show, in the rotating frame the interaction Hamiltonian oscillates rapidly (on the order of the Larmor frequency) about a slowly varying average value.
If we choose our simulation time steps $dt$ be sufficiently small, the slowly varying magnitude of the interaction can be treated as constant on any given time step, and we can use this average Hamiltonian theory to eliminate the high frequency terms entirely.
In order to characterize the ``slowly varying'' and ``rapidly varying'' parts of the interaction and set limits on our choice of $dt$, we first examine how the interaction term behaves in the rotating frame. Specifically, the high frequency terms will disappear from the effective Hamiltonian, allowing us to use a much larger time step than what would be required for a typical Larmor frequency in the tens of MHz.
Next,  we will   consider  different cases of $d=x,y,z$ in \mbox{Equation \ref{eq:rot_ref1}}.

The case $d = z$ in Equation \ref{eq:rot_ref1} is uninteresting, as $I_z$ commutes with $U$.  In this case, the choice of $dt$ depends only on the magnitude of $\alpha_z$.  For $d = x, y$; $U$ has the effect of rotating the spin operator about the $z$ axis by an amount $\nu t$:
\begin{align}
    \label{eq:rotating_spin_ops}
    \begin{split}
        U^\dag I_x U & = I_x \cos \nu t - I_y \sin \nu t\\ 
        U^\dag I_y U & = I_y \cos \nu t + I_x \sin \nu t \, .
    \end{split}
\end{align}
By taking advantage of the cyclic property of the trace, $M^{(i)}_d$ can also be rewritten relative to the rotating density matrix $\rho_R$:
\begin{align}
    \label{eq:rotating_m}
    M^{(i)}_d = \sum_{j \neq i} f(r_{ij}) \langle U^\dag I^{(j)}_d U \rangle_R \, 
\end{align}
where $r_{ij} \equiv |\vec{r}_i - \vec{r}_j|$ and $\langle I_d^{(j)} \rangle_R$ indicates the expectation value taken with respect to $\rho^{(j)}_R$.  For simplicity, moving forward we define $\langle - \rangle \equiv \langle - \rangle_R$, unless otherwise noted. Returning to the Hamiltonian, by combining the expressions in Equations \ref{eq:rotating_spin_ops} and \ref{eq:rotating_m} and separating the oscillatory (those with an explicit $\sin n\nu t$ or $\cos n \nu t$) and non-oscillatory terms, our Hamiltonian in the rotating frame becomes
\begin{equation}
    \begin{aligned}
        H^{(i)}_R = & -(\Delta \nu + a^{(i)}_z) I_z^{(i)} \\ & - (a^{(i)}_{x+} + a^{(i)}_{x-} \cos 2\nu t - a^{(i)}_{y-} \sin 2\nu t) I_x^{(i)}\\
        & - (a^{(i)}_{y+} - a^{(i)}_{y-} \cos 2\nu t - a^{(i)}_{x-} \sin 2\nu t) I_y^{(i)}
    \end{aligned}
\end{equation}
with coefficients $a_i$ defined by
\begin{equation}
    \begin{gathered}
    a^{(i)}_z = \gamma \alpha_z \sum_{j \neq i} f(r_{ij}) \langle I_z^{(j)} \rangle \\
    a^{(i)}_{\pm} = \frac{\gamma}{2}(\alpha_x \pm \alpha_y) \sum_{j \neq i} f(r_{ij}) \langle I_d^{(j)} \rangle \, .
    \end{gathered}
\end{equation}
We define an in-plane interaction strength as $\alpha \equiv \alpha_x + \alpha_y$, an anisotropy factor $\eta$ as
\begin{align}
    \label{eq:anisotropy_def}
    \eta \equiv \frac{\alpha_x - \alpha_y}{\alpha} \, ,
\end{align}
and effective frequencies $\omega^{(i)}_d$ as
\begin{align}
    \label{eq:omega_xy_def}
    \omega^{(i)}_d \equiv \sum_{j \neq i} f(r_{ij}) \langle I_d^{(j)} \rangle \, .
\end{align}
Finally, we define $\omega_\perp^{(i)} \equiv \omega_x^{(i)} + i\omega_y^{(i)}$.  With these substitutions, the Hamiltonian becomes
\begin{equation}
    \label{eq:hamiltonian_anisotropy}
    \begin{aligned}
        H_R^{(i)} & = - \left(\Delta \nu + a_z^{(i)}\right)I_z^{(i)} \\
        & - \frac{\gamma \alpha}{4}\left( \omega_\perp^{(i)} \left( 1 + \eta e^{2i\nu t} \right) + \bar{\omega}_\perp^{(i)} \left( 1 + \eta e^{-2i\nu t} \right) \right) I_x^{(i)} \\
        & - \frac{i\gamma \alpha}{4}\left( \bar{\omega}_\perp^{(i)} \left( 1 - \eta e^{-2i\nu t} \right) - \omega_\perp^{(i)} \left( 1 - \eta e^{2i\nu t} \right) \right) I_y^{(i)}
    \end{aligned}
\end{equation}
where $\bar{\omega}_\perp$ is the complex conjugate of $\omega_\perp$.
We examine the effect of the $\eta$ terms using the Magnus expansion \cite{BLANES2009151,PhysRevA.41.2311,MANANGA20161}. We find that the $\eta$ terms can generally be ignored when the ratio of the interaction strength to the Larmor frequency is sufficiently small.

\subsection{Effective Time Scale $\Omega$}
\label{sec:methods_effective_scales}
The Magnus expansion is only valid if the norm of the Hamiltonian is much smaller than the time interval over which the expansion is calculated.  One such norm is given by \cite{https://doi.org/10.1002/cmr.a.21414}
\begin{align}
     |\mathbf{\Omega}| \equiv \Omega = \bigg| \textrm{Tr}\left( \left[ H_R^{(i)} \right]^2 \right) \bigg|^{\nicefrac{1}{2}} \ll \nicefrac{1}{dt} \, .
\end{align}
This condition can also be used as a limit on the validity of our assumption that the coefficients $\omega_d^{(d)}$ are constant over the period $dt$ of the Magnus expansion, (the time step of the simulation).  For an isotropic planar interaction, we find
\begin{align}
\label{eq:Omega}
    \Omega = \frac{1}{\sqrt{2}}\sqrt{(\Delta \nu + \gamma \alpha_z \omega_z^{(i)})^2 + (\nicefrac{\gamma \alpha |\omega_\perp|}{2})^2 }.
\end{align}
We now derive an upper bound on the value of $\Omega$ to estimate the minimal time step necessary for accurate time propagation.
To do so, we will replace the dynamic or site-dependent variables in Eq.~\ref{eq:Omega} with estimated or maximal values.
As each individual spin will have a different value of the parameter $\Delta \nu$, we substitute $\Delta \nu$ with an ``average'' value, \ie the linewidth $\Gamma$.
We additionally define quantities $w$ and $w_z$ to simplify the local magnetization paramaters $\omega_z^{(i)}$
\begin{equation}
    \label{eq:omega_def}
    \begin{gathered}
        w \equiv (\alpha_x + \alpha_y) \sum_{j \neq 1} f(r_{1j}) \\
        w_z \equiv \alpha_z \sum_{j \neq 1} f(r_{1j}) \, .
    \end{gathered}
\end{equation}
These two terms, which we call the planar and out-of-plane ``interaction weight'' respectively, represent the largest possible value of the mean-field magnetization.
We also need to simplify the $\braket{I_d^{(j)}}$ term in Eq.~\ref{eq:omega_xy_def}, which we will set to their maximal expectation value, $\braket{I_d^{(j)}} = S = 1/2$.
With these substitutions, we arrive at
\begin{align}
    \label{eq:big_omega_def}
    \Omega = \sqrt{\nicefrac{(\Gamma + \gamma s w_z)^2}{2} + \nicefrac{(\gamma sw)^2}{8}}
\end{align}
which defines the maximal possible strength of the interaction.  The condition for our approximation of the time evolution to be valid is given by
\begin{align}
    \label{eq:slowly_varying_condition}
    \Omega \ll \nicefrac{1}{dt} \, .
\end{align}
In the case of an isotropic planar interaction, our Hamiltonian reduces to one nearly identical to the original Hamiltonian in the laboratory frame, but with a shifted Larmor frequency:
\begin{align}
    H^{(i)}_R = & -\left( \nu_i - \nu \right) I_z^{(i)} - \gamma \sum_{d = x,y,z} \alpha_d M_{d,L}^{(i)} I_d^{(i)} \, ,
\end{align}
where we require $\alpha_x = \alpha_y$ and we have defined the ``local magnetization''
\begin{gather}
    \label{eq:local_m_def}
    M_{d,L}^{(i)} = \sum_{j \neq i}f(r_{ij})\langle I_d^{(j)} \rangle_R \, \quad \textrm{for \, $d = x, y, z, +$}\,  
\end{gather}
where $d = +$ corresponds to the raising operator $I_+$, which corresponds to the complex valued planar magnetization.  This parameter is equivalent to our parameters $\omega_d^{(i)}$ introduced in Eq.~\ref{eq:omega_xy_def}, but now in the rotating frame. 
But even for an anisotropic interaction, the behavior of the $x$ and $y$ axes are not truly distinct.
The $M_{x}$ and $M_{y}$ terms combine to form one effective planar interaction term, whose strength is the average of the individual strengths, ($\alpha_x$ + $\alpha_y$)/2.  For a spin $\nicefrac{1}{2}$ nuclei, this planar averaging effect is evident in the simplified form of the rotating-frame Hamiltonian:
\begin{gather}
    H^{(i)}_R = \frac{\hbar}{2}\begin{pmatrix}
    -\Delta \nu - \gamma \alpha_z M_{z,L}^{(i)} & -\gamma \frac{\alpha}{2} M_{+,L}^{(i)} \\
    -\gamma \frac{\alpha}{2} \overline{M}_{+,L}^{(i)} & \Delta \nu + \gamma \alpha_z M_{z,L}^{(i)} 
    \end{pmatrix}
\end{gather}
for $\overline{M}$ the complex conjugate of $M$ and $\alpha \equiv \alpha_x + \alpha_y$.

\subsection{Liouville Space}
\label{sec:liouville}
We model our material as an open quantum system in Liouville space, which allows us to introduce dissipation to an external bath (usually assumed to be a combination of electrons and phonons).  In Liouville space, the $d \times d$ density matrix is flattened into a $d^2 \times 1$ vector; in our simulation, we use row major order to make this conversion.  On account of this convention, it can quickly be shown that left and right multiplication by some operator $\mathcal{O}$ are written
\begin{equation}
    \begin{gathered}
    \mathcal{O}\rho \longrightarrow (\mathcal{O} \otimes \mathds{I})\rho_L \\
    \rho \mathcal{O} \longrightarrow (\mathds{I} \otimes \mathcal{O}^T)\rho_L \, .
    \end{gathered}
\end{equation}
With this convention in mind, the time evolution equation for our density matrix becomes
\begin{equation}
\label{eq:linblad_eom}
    \begin{gathered}
    \frac{d\rho^{(i)}_R}{dt} = \frac{i}{\hbar}[\rho^{(i)}, H^{(i)}_R] \longrightarrow \frac{d\rho^{(i)}_{R,L}}{dt} = \frac{i}{\hbar}\hat{L}_R^{(i)}\rho^{(i)}_{R,L} \\
    \textrm{with} \\
    \hat{L}_R^{(i)} = \mathds{I} \otimes \overline{H}_R^{(i)} - H_R^{(i)} \otimes \mathds{I} \, .
    \end{gathered}
\end{equation}

For a time independent Hamiltonian, the above equations have the solution
\begin{align}
    \rho_{R,L}^{(i)}(t) = e^{i \hat{L}_R^{(i)} t/\hbar}\rho_{R,L}^{(i)}(0).
\end{align}
Our Hamiltonian, which depends on the expectation values of the spins in the ensemble, is not time independent.  However, as previously established, we can approximate the Hamiltonian as time independent on time steps $dt$ so long as the condition in Equation \ref{eq:slowly_varying_condition} is met.  Explicitly, we time evolve our density matrices in steps $dt$
\begin{align}
    \rho_{R,L}^{(i)}(t + dt) \approx e^{i \hat{L}_R^{(i)}(t) \, dt/\hbar}\rho_{R,L}^{(i)}(t).
\end{align}
    
In an experimental setting typical of the systems we are modelling, the so called spin-lattice relaxation time $T_1$ is often much slower than $\tau$ and can be generally be ignored for a single spin echo measurement.  However, the spin-spin relaxation time $T_2$ is often similar in scale to $\tau$.
Although our model already included the mean-field effect of the spin-spin interaction, our approximation is not designed to capture the spin decoherence caused by local $J$ terms along chemical bonds.
To that end, we include the $T_2$ effect in our simulation by considering additional non-Unitary Linblad jump operators $L_n$ via the following master equation~\cite{AmShallem2015}:
\begin{align}
\label{LimNop}
    \frac{d\rho}{dt} = -\frac{i}{\hbar}[H,\rho] & + \sum_{n} \gamma_n \left( L_n \rho L_n^\dag - \frac{1}{2} \left\{L^\dag_n L_n, \rho \right\} \right) \, ,
\end{align}
where $\gamma_n$ is the dissipation rate for operator $n$ and $\{A,B\} = AB + BA$ is the anti-commutator.
We note that the commutator of $\rho$ and $H$ simply reproduces Eq. \ref{eq:linblad_eom}, while the remaining terms are introduced to capture how non-unitary (dissipative) operators act upon the density matrix. Although we are primarily concerned with decoherence via $T_2$ processes (\ie $L_n = I_z$), we can easily include terms that account for emission ($L_n = I_-$) and absorption ($L_n = I_+$), which simulate $T_1$ and thermal effects.  
 We emphasize that  the purpose of  the Lindblad terms introduced here is to capture the effects of ``long-range'' interactions (mediated through e.g. the electron gas). These Lindblad terms 
do not account for other sources of  magnetic noise, e.g. local field differences due to other, shorter-range magnetic interactions.  
 For example, if there was another magnetic nuclei in the material that was not coupled
 to the electrons as strongly as the primary species, then this would be a strong source 
 of ``$T_2$'' like decay, leading each ``primary'' nuclei to see a slightly different  magnetic field.  
 Dissipation rates for operator $n$, in \mbox{Eq. \ref{LimNop}},  are parameters of the model. The simulation
 does not assume a particular environment that would inform the choice of $\gamma_n$.  Therefore, they would be set by (or fit to) the dissipative effects of some desired environment. All $n$
 are not the same. The different $\gamma_n$ represent the effects of different modes of
  dissipation via the various jump operators (e.g. $\gamma_\pm$ for emission/absorption, respectively.) 
 and are not necessarily equal. They can be set independently in the simulation.

Taking into account our row-major ordering and the self-adjoint nature of the spin operators, the inclusion of dissipation results in an additional term on the right hand side of the equation of motion (Eq. \ref{eq:linblad_eom})
\begin{align}
    \hat{J} = \sum_{d = z, +, -} \frac{\Gamma_d}{\hbar^2} \left( -\frac{1}{2} \bigg[ I_d^2 \otimes \mathds{I} + \mathds{I}\otimes \overline{I}_d^2 \bigg] + I_d \otimes \overline{I}_d\right)
\end{align}
and yields the following equation for the time evolution step in our simulation:
\begin{align}
    \rho_{R,L}^{(i)}(t + dt) = e^{\left( i\hat{L}_R^{(i)}(t)/\hbar + \hat{J} \right)dt} \rho_{R,L}^{(i)}(t).
\end{align}

With the dissipation included, we have to re-examine the Magnus expansion.  The terms calculated in the Hilbert space representation will still be valid, but there will be additional terms which represent the mixing of the dissipation and the interaction.  As before, the first order term is simply the time-averaged term and is equivalent to the isotropic case.  The second order term is identical to the second order term from the Hilbert space representation, plus an additional term that accounts for the mixing of the dissipation.  This additional term has factors that look like
\begin{align}
    \label{eq:dis_magnus_ex}
    \frac{\gamma \alpha \omega \eta}{\nu}\sum_{i = 1}^{3} \beta_i \Gamma_i
\end{align}
where the $\beta_i$ are either $\pm1$ or $\pm 3$ depending on the term and $i$.  We see the exact same scaling as in the previous examination, as well as a mixing of the dissipation and interaction terms.  Given the dramatic increase in complexity of the calculation in Liouville space, we only calculate the first and second order terms.

Although we do not analyze the behavior of the model in the presence of dissipative effects in this discussion, the nature of the simulation requires many of the desired features to be hard-coded.  We therefore structure the simulation to give the option for dissipation in order to enable future exploration of the behavior of the model.

\section{RESULTS}
\label{sec:results}

We now examine the time dependence of the total magnetization in the presence long-range spin-spin coupling.
As mentioned in Section~\ref{sec:methods_ensemble}, the rotating frame Hamiltonian has an overall scaling which can be removed if one also rescales the units of time.
To provide results for a general material, we therefore work in units of the magnetic linewidth $\Gamma$ and the characteristic magnetic dephasing time $1/\Gamma$.

We first simulate the spin dynamics for an infinite-range interaction.
Under this interaction, all spins experience a time-dependent force which is proportional to the net magnetization of the material.
This is equivalent to a shorter-range interaction but in the limit of an infinite number of nearest neighbors (infinite bond dimension).
Under this simplifying assumption, we study the dependence of the spin-echo and the FID on interaction magnitude and the applied pulse strength.

\begin{figure*}[t]
    \centering
    \includegraphics[width=0.55\linewidth]{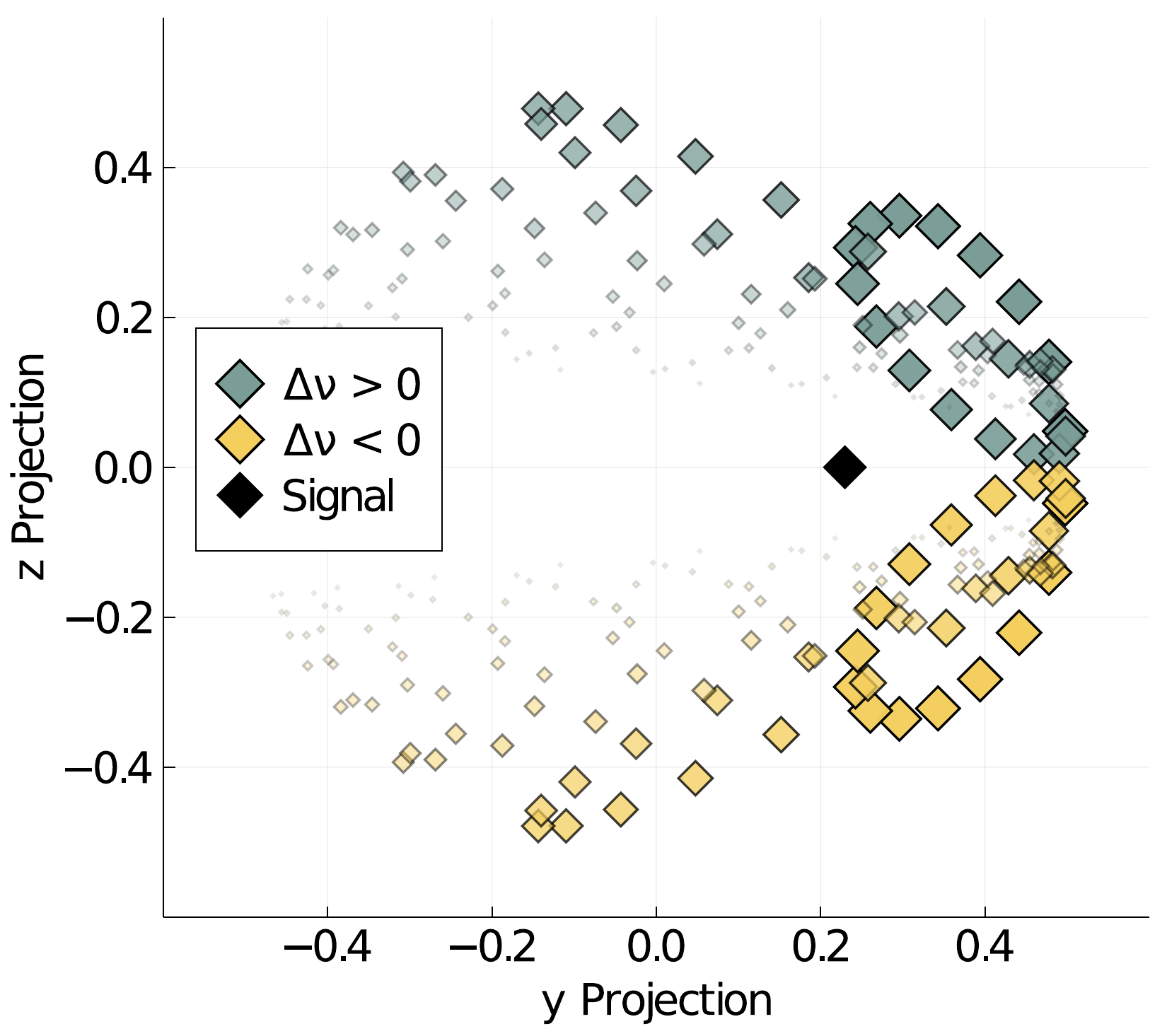}
    \caption{$\langle I_z \rangle$ vs $\langle I_y \rangle$ for each nuclear spin at $t = \nicefrac{2}{\Gamma}$ for an echo with $\tau = \nicefrac{2.5}{\Gamma}$ and $T_2^* = \nicefrac{1}{\Gamma}$.
    Each marker indicates the $(y,z)$ components of a single spin.
    To generate this figure we sample the spins' frequency distribution uniformly, but each spin's effective contribution to the mean field is proportional to the distribution's probability at that frequency.
    Marker transparency and size follows the distribution's probability at each spin's $\Delta \nu$, with larger and more opaque markers indicating higher effective contribution.}
    \label{fig:tilted_axis_spins}
\end{figure*}

After gaining insight into the effect of the infinite range coupling, we will see what stays the same and what changes when reducing the range of the interaction.
The simulations with finite-ranged interactions can be compared to those with infinite-range interactions if they have similar average interaction weights $\omega$, as defined in Section~\ref{sec:methods_effective_scales}.
We find a smooth crossover between the behavior at infinite range and the well-understood behavior at short ranges (magnetic decoherence).

\subsection{Global Interaction}
\label{sec:global_theory}

\begin{figure*}[t]
    \centering
    \includegraphics[width=0.55\linewidth]{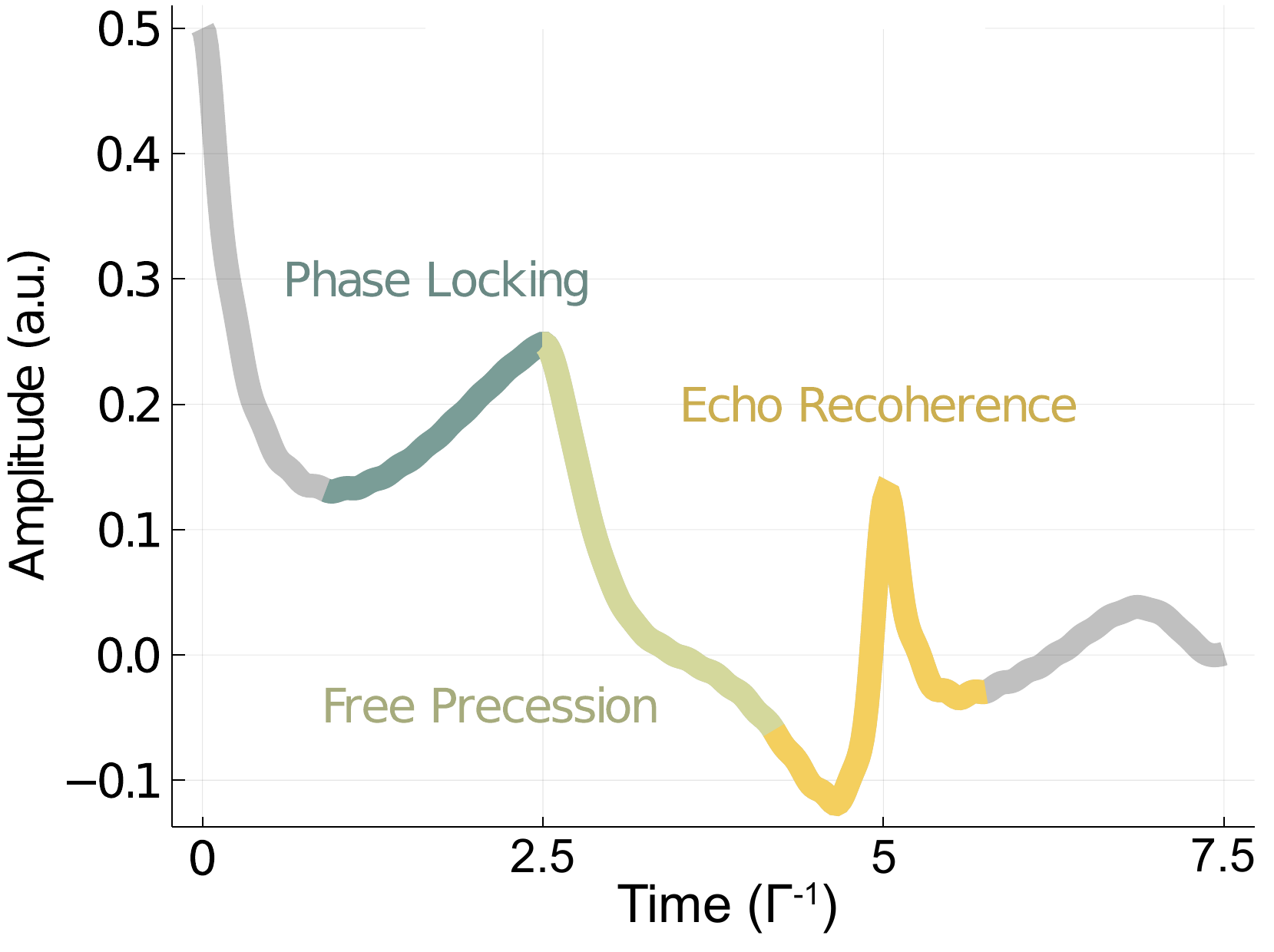}
    \caption{Echo phases labeled by color for a simulation with total planar interaction weight $\omega = 3.05\Gamma$.  The $\pi$ pulse occurs at $\tau = 2.5\Gamma^{-1}$, causing an echo at $2\tau = 5\Gamma^{-1}$.}
    \label{fig:spin_axis_angle_vs_dv}
\end{figure*}

\subsubsection{Behavior of individual spinors}
The $I_x, I_y$ terms of the spin-spin interaction can induce rotations in addition to the precession caused by the Zeeman term.
Just after they are tipped into the plane, the planar interactions $\alpha_x=\alpha_y$ can combine with the small, non-zero $x$ component of the spin in the rotating frame and apply an in-plane torque.
This causes additional rotation of the spins out of the $x-y$ plane.
This torque is either ``up'' (towards the $+z$ axis) if $\Delta \nu > 0$ or ``down'' (towards the $-z$ axis) if $\Delta \nu < 0$, as the sign of deviation from the rotating frame determines the sign of the local $\braket{I_x}$ and thus the sign of the torque.
In other words, the precession axis of each spin slightly tilts away from the $z$ axis and towards the $\pm y$ axis.
The tilting of the precession axes can also shift the average precession frequency for the ensemble.
That is to say, the mean field magnetization manifests as a Knight shift of the spectral line.

This effect is captured in Figure~\ref{fig:tilted_axis_spins}, where we have used a weighted sampling (as opposed to random sampling) of the spin distribution to allow for easy interpretation of how the center of the distribution behaves.
Because the spins now precess about a tilted axis, they do not fully dephase and a lingering net magnetization will remain along the $y$ axis.
We call this a ``phase locking'' effect, and its continuation of transverse magnetization out to the $\pi$ pulse at $t=\tau$ is a primary driver of the multi-peak behavior we observe in the simulated echoes.
For a sufficiently strong interaction, this phase locking effect can result in spin coherence times much longer than the echo time.
The phase locking caused by the planar interaction can be interpreted as an internal version of spin-locking, which is the extension of transverse magnetization during an echo via externally applied pulses.

\begin{figure*}[t]
    \centering
    \includegraphics[width=\linewidth]{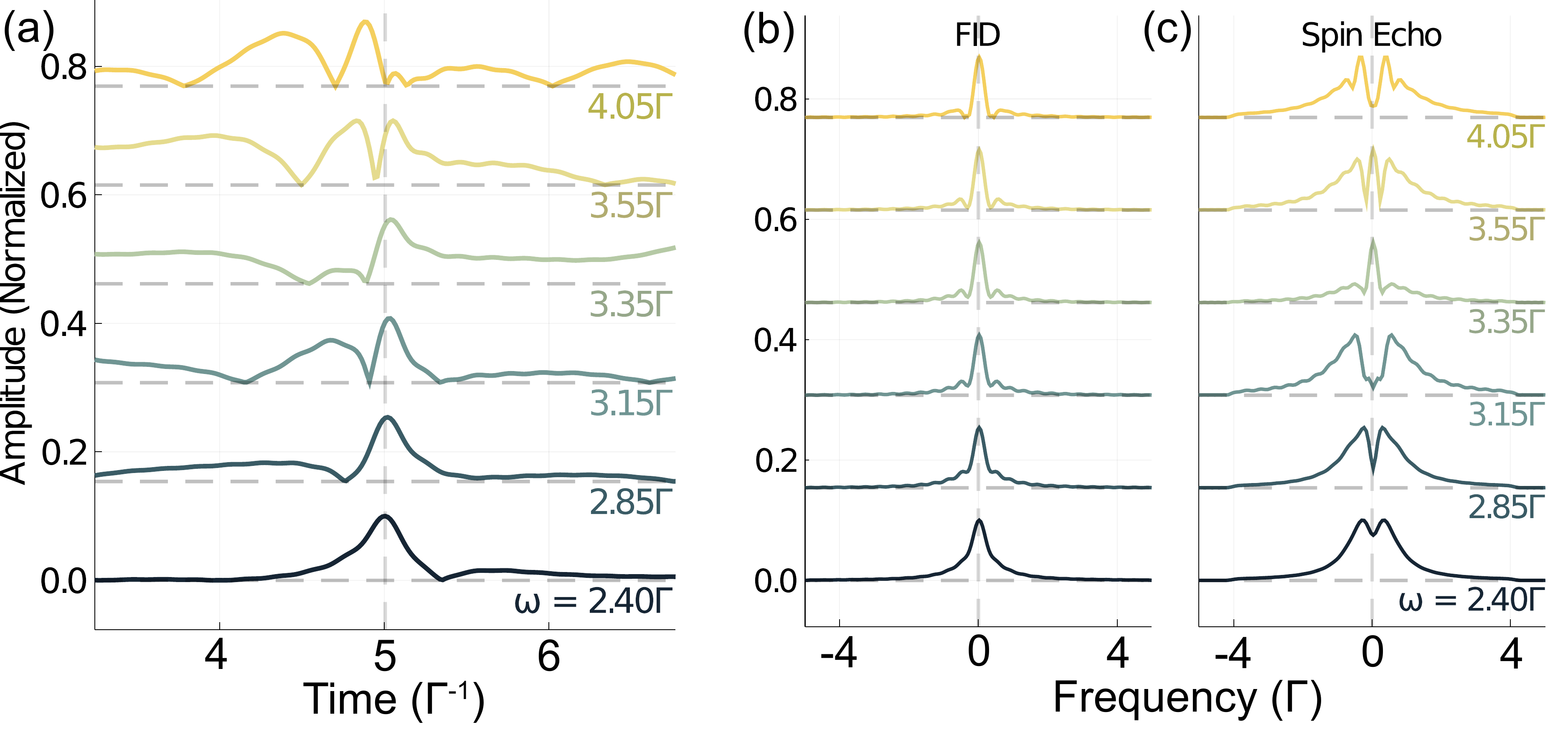}
    \caption{(a) Spin echoes for a global interaction with varying planar weight $\omega$. The $\omega$ value for each simulation is given on the right of each curve. (b) Fourier transform (spectrum) of the FID region ($t < \tau$). (c) Fourier transform (spectrum) of a region centered on the echoes. }
    \label{fig:echo_sample_long_xi}
\end{figure*}

\subsubsection{The interacting spin-echo}

In Figure \ref{fig:spin_axis_angle_vs_dv}, we show the typical time evolution of an interacting spin ensemble under the spin echo protocol.
The phase locking effect is seen most prominently during the free induction decay (FID), which occurs just after the $\nicefrac{\pi}{2}$ pulse ($t=0$).  During the FID, the net magnetization is very strong and the spins are beginning to dephase about the $z$ axis.  Unlike in an interaction-free FID, the additional rotation about the $y$ axis causes the spins to spiral out, gaining a non-zero $z$ component as they rotate.  Eventually, as the net magnetization decreases and the spins dephase, the precession axis for the spins partially stabilize, resulting in a nutation effect as the axes oscillate in some slice of the $y-z$ plane.  This effect is observed in the time domain of the FID as a persistent signal which oscillates about some non-zero average value.  In the frequency domain, the interaction's rotation opposes the Larmor precession of the spins by an amount proportional to the value of $\langle I_z \rangle$ for the spin.

The $\pi$ pulse ($t = 50$ $\mu$s) then flips the spins about the $y$ axis, causing the $z$ component of each spin to change sign. Now the interaction no longer opposes the natural precession of the spins but reinforces it.  This results in the spins rapidly dephasing and a sudden drop in the signal.  As they precess and the signal decays, the interaction disappears and the spins enter a period of free precession (as the mean field magnetization is zero).  However, the lingering coherence from the phase locking as well as the force of the interaction causes the positive ($\Delta \nu > 0$) and negative ($\Delta \nu < 0$) spins to precess in ``clumps''.  
Depending on the strength of the interaction, the degree of coherence from the phase locking, and the orientation of the spins at the $\pi$ pulse, this ``clumping'' effect can result in the spins re-cohering along the $-y$ axis at slightly different times, creating the multiple peaks we see prior to the echo.
Once the spins precess and re-cohere along the $-y$ axis, a second period of phase locking can emerge as the alignment of the spins and the interaction results in the interaction once again opposing the natural precession of the spins.

As we approach $t = 2\tau$, the spins begin to realign due to the reversal of the Larmor precession induced by the $\pi$ pulse.
This re-coherence seems to favors alignment with the $+y$ axis, causing the coherence along the $-y$ axis to suddenly and rapidly decay and the net magnetization to rapidly change sign.
With a few exceptions, peaks emerging after the echo are typically highly attenuated and negligible compared to the echo and pre-echo peaks.

\subsubsection{Dependence on interaction strength}
In Figure \ref{fig:echo_sample_long_xi}(a) we plot the results for simulations with varying interaction weights, given in units of the linewidth $\Gamma$.
As the interaction strength increases, the net magnetization caused by the phase locking during the FID grows, causing larger peaks prior to the echo. At very stronger interaction strengths, these effects eventually destroy the spin echo entirely ($\omega > 4 \Gamma$).  Beyond this regime, the phase locking begins to dominate, and ultimately the system enters a regime wherein the $\Delta \nu$ precession plays almost no role in the dynamics and interaction-driven phase locking makes up the entirety of the simulation.

The spectra of these simulations are shown in Figure~\ref{fig:echo_sample_long_xi}(b,c).
There are two ways the interaction's effect is exhibited in the spectral domain. 
The first is the previously mentioned narrowing of the spectrum during the FID, caused by the adversarial effect of the interaction relative to the $\Delta \nu$ precession (Figure~\ref{fig:echo_sample_long_xi}(b)).  The other observed feature is an effect akin to hole burning, which is a disappearance of spectral weight only at the center of a spectral line (Figure~\ref{fig:echo_sample_long_xi}(c)).    The hole burning effect is driven by the change in the sign of the interaction at the spin echo. 
The refocusing of the spins caused by the $\pi$ pulse pulls the net magnetization of the ensemble towards the axis of the initial signal during the FID. 
This change in the sign of the interaction causes the interaction to augment the natural precession of the spins, but on account of the frequency-dependent behavior of the spins it affects those spins near $\Delta \nu = 0$ most strongly. 
The spins with frequencies near $\Delta \nu = 0$ are therefore pulled strongly away from $\Delta \nu = 0$, resulting in a hole burning effect.

\begin{figure}[t]
    \centering
    \includegraphics[width=\linewidth]{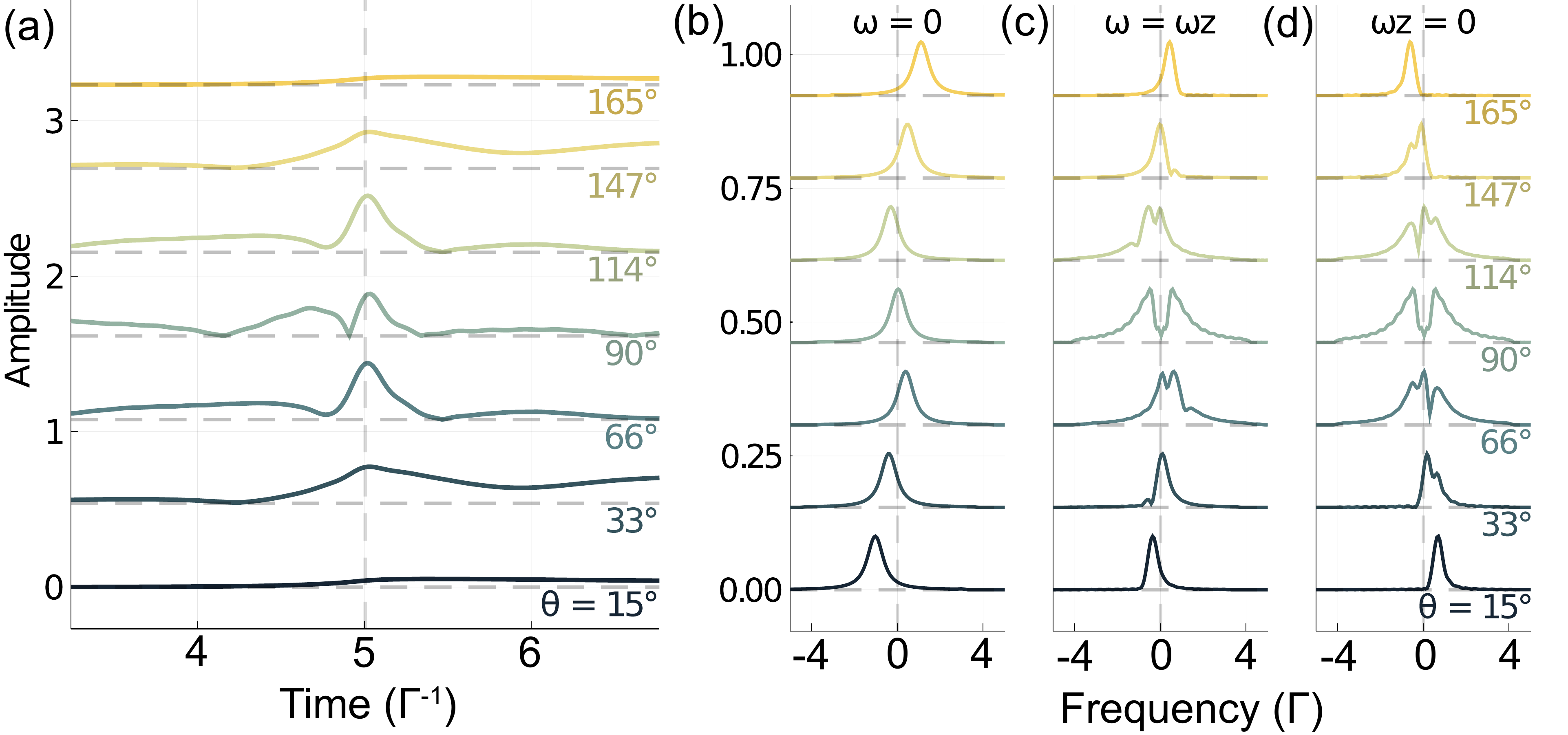}
    \caption{
    (a) Spin echoes for a global interaction with $\omega = 3.15\Gamma$ but varying flip angles $\theta$. (b-d) Spectra dependence on flip angle $\theta$ for interaction $(\omega, \omega_z)$ equal to (b) $(0, 3.15\Gamma)$, (c) $(3.15\Gamma, 3.15\Gamma)$, and (d) $(3.15\Gamma, 0)$.}
    \label{fig:global_vs_pulse_echoes}
\end{figure}

The introduction of a $z$-axis spin interaction ($\alpha_z$) in the global interaction has no effect if the spins are flipped perfectly into the plane. 
Given the interaction is global, the $z$ coupling will only impact the dynamics of the spins when there exists a $z$-component to the net magnetization in the $z$ component.
Therefore, we now turn to the tip angle dependence of the spin dynamics.

\begin{figure}[t]
    \centering
    \includegraphics[width=\linewidth]{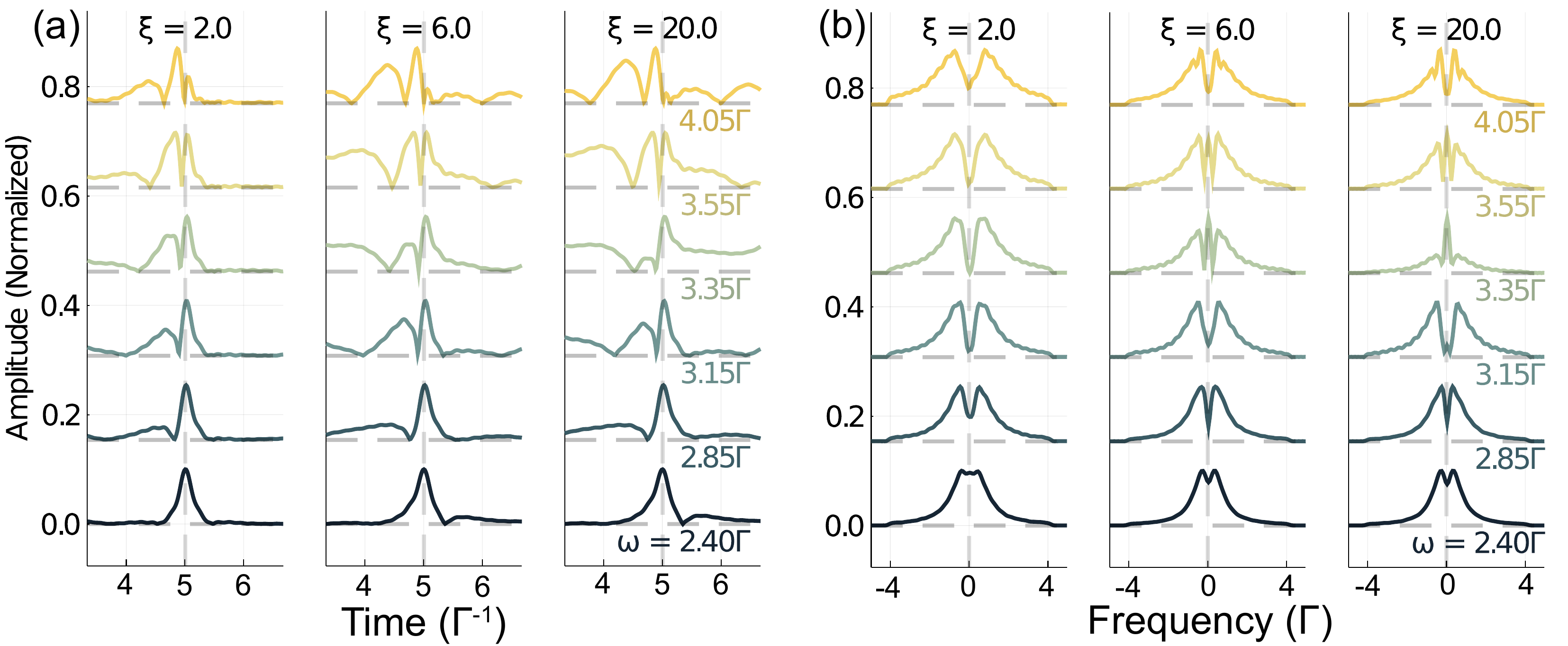}
    \caption{(a) Spin echoes for a local interaction with $\xi \in [2,6,20]$. The planar interaction weight $\omega$ is given on the right of each set of curves, and $\omega_z = 0$. (b) Spectra for the same echoes.}
    \label{fig:p_vs_xi_echo_comparison}
\end{figure}

\subsubsection{Pulse angle dependence}
In Figure \ref{fig:global_vs_pulse_echoes}(a), we show time-domain simulations with a selection of pulse angles $\theta$ from $15^\circ$ to $165^\circ$.
For pulse angles $\theta$ in the range $[60^\circ,120^\circ]$, any deviation from $90^\circ$ dampens the multi-peak behavior on account of the effective reduction of the interaction strength.
For $\theta$ outside this range, the multi-peak behavior vanishes almost entirely, and a post-echo phase locking effect is observed instead. 
This region of phase locking is caused by the additional global $z$ magnetization, which applies a consistent torque to the spins and lowers the necessary interaction threshold for the phase locking to occur. 
For $\theta < 30^\circ$ or $\theta > 150^\circ$ the effect of the interaction on the time domain echo begins to disappear entirely as the driving force of the temporal behavior, the planar interaction, is attenuated by the reduced in-plane projection of the ensemble magnetization. 
Changing the $z$ coupling has little effect on this behavior except to dampen the phase locking effects at $\theta \approx 30^\circ, 150^\circ$.

In Figure~\ref{fig:global_vs_pulse_echoes}(b-d) we plot the spectral transform of spin echoes from similar simulations, including the relative strength of planar vs out-of-plane interaction effects.
Most clearly, the $z$ interaction causes an overall pulse dependent frequency shift (Figure~\ref{fig:global_vs_pulse_echoes}(b))
When the spins are not flipped fully into the plane, or are flipped past it, the consistent non-zero $\langle I_z \rangle$ of the spins breaks the symmetry between the spins of different sign $\Delta \nu$ and the spin ensemble experiences a net torque, resulting in an overall frequency shift. 
The effect from $\omega$ and $\omega_z$ oppose each other, with the $\omega_z$ shifts slightly dominating when $\omega = \omega_z$ (Figure~\ref{fig:global_vs_pulse_echoes}(c)).  The exact form of the dependence of these shifts on the flip angle in the global case is derived in our companion work \cite{theory_paper}.
There is a smooth crossover between an effective Knight shift and hole burning as one moves the pulse angle towards $90^{\circ}$ from either direction (Figure~\ref{fig:global_vs_pulse_echoes}(d))
\begin{figure}[!tbp]
    \centering
    \includegraphics[width=0.57\linewidth]{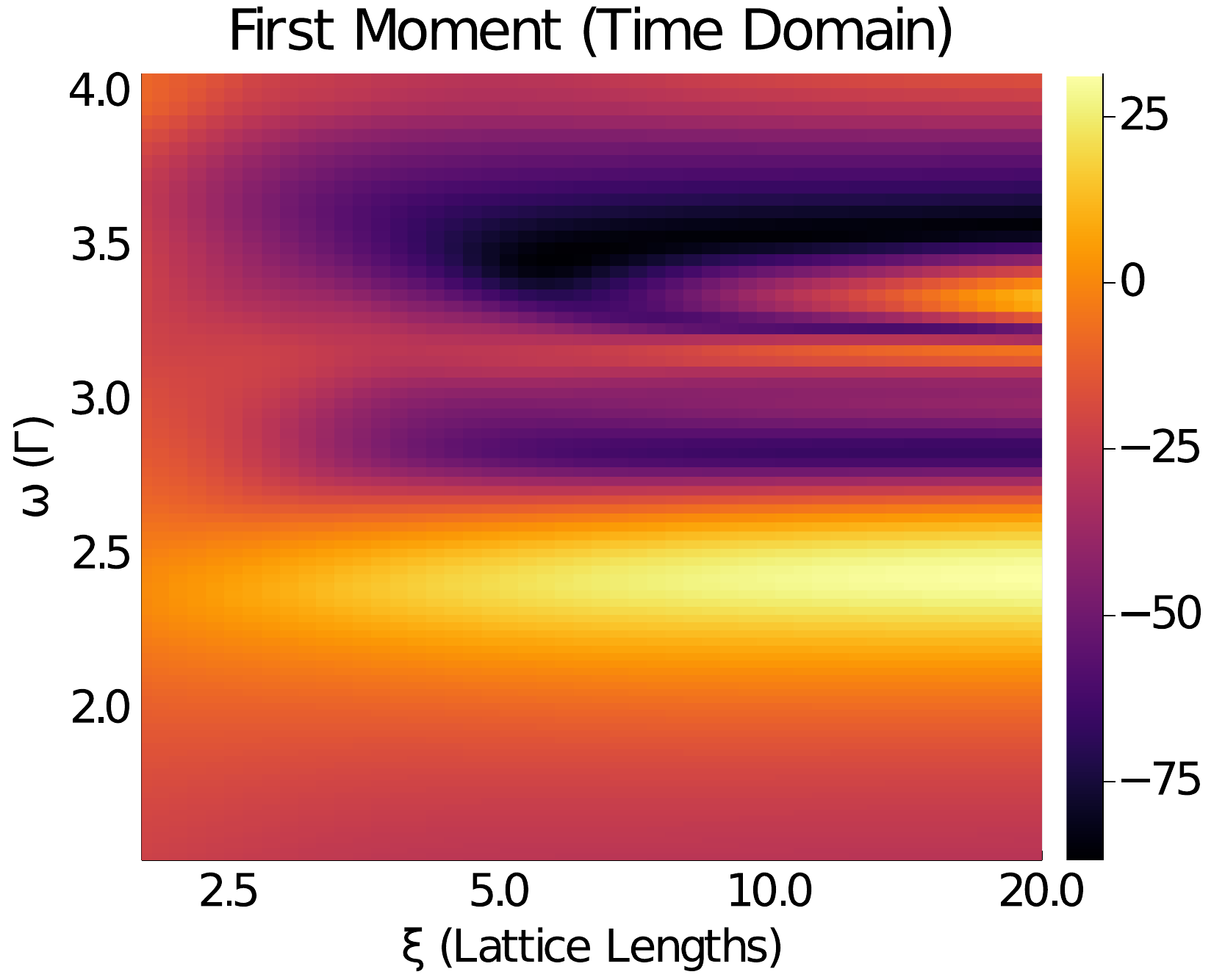}
    \caption{First moment $\mu_1$ of the spin echo as defined in Eq.~\ref{eq:first_moment_def}, with arbitrary units.
    The simulated echoes had parameters ranging from $\omega = 1.5\Gamma$ up to $\omega = 4.05\Gamma$ (y-axis) and $\xi = 2.0$ to $\xi = 20.0$ (x-axis), with $\omega_z = 0$ for all.
    A large negative first moment was usually caused by a pre-echo peak, while a positive first moment indicates post-echo ringing in the signal.}
    \label{fig:first_moment_p_vs_xi}
\end{figure}

\subsection{Local Interaction}
\label{sec:local_interaction_results}

We now consider interactions governed by Eq. \ref{eq:f_gaussian}, namely a gaussian interaction between spins with range given by $\xi$.
Figure \ref{fig:p_vs_xi_echo_comparison} displays the variety of echo behaviors for several $\xi$ values for in-plane coupling only ($\alpha_z = 0$).
The echo amplitudes have been normalized to accentuate the different behaviors, but the stronger $\omega$ echoes have generally smaller amplitudes.  These echoes were run on a $400 \times 400$ lattice with $n = 160,000$.
For a very short range interaction between nuclei ($\xi \leq 2$), the dynamics are dependent on local fluctuations in frequency and magnetization.  Although phase locking is still observed, the degree of coherence is limited by local fluctuations, which ultimately dampens the effect and attenuate additional peaks.  Echoes from an intermediate interaction range ($\xi = 6$) typically behave similarly to those of nearly-global interactions ($\xi = 20$), with the notable exception of the post-echo phase locking mentioned previously.
\begin{figure}[!tbp]
    \centering
    \includegraphics[width=\linewidth]{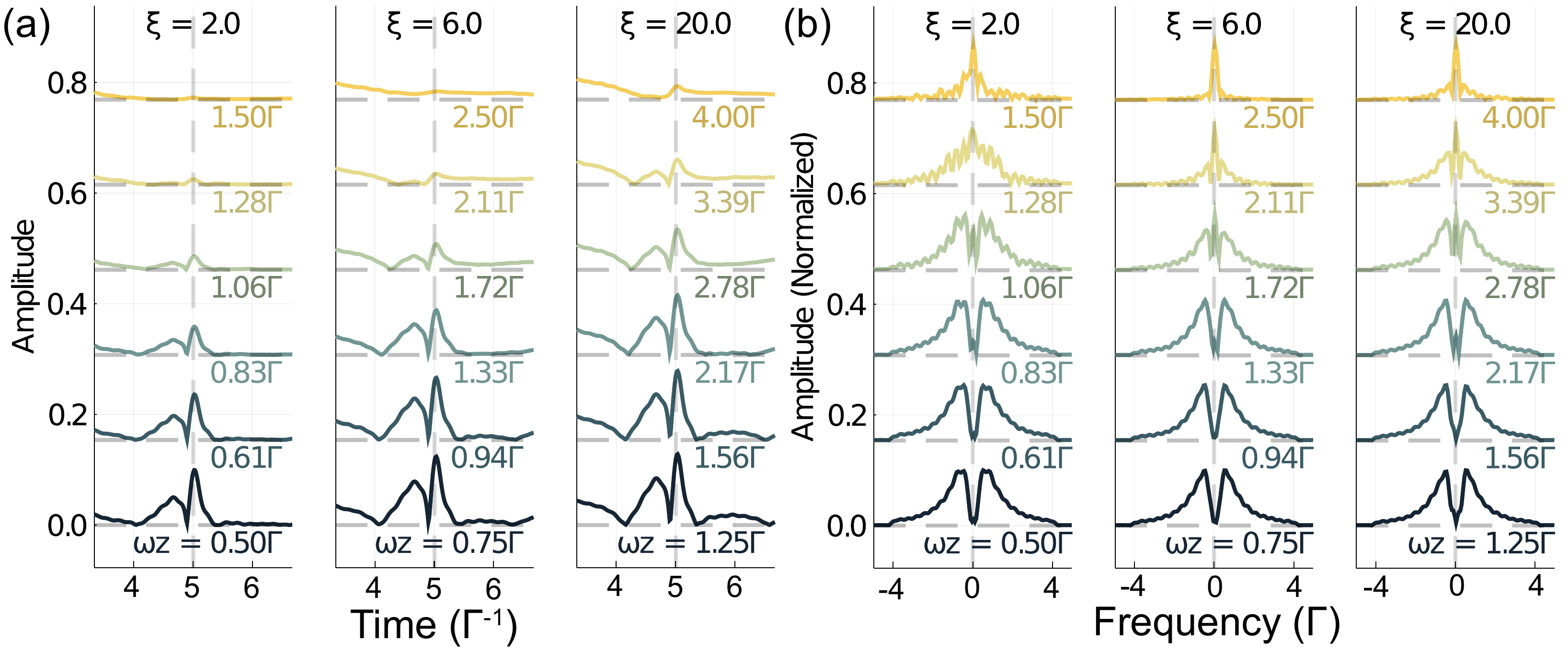}
    \caption{(a) Spin echoes for a local interaction with $\xi \in [2,6,20]$. The out-of-plane interaction weight $\omega_z$ is given on the right of each curve, and $\omega = 3.15 \Gamma$. Note that the amplitudes have not been normalized to emphasize the loss of magnetization at large $\omega_z$. (b) Spectra for the same echoes, but with normalized signal.}
    \label{fig:echo_compare_all_xi_with_z}
\end{figure}
We can see the continuous dependence of the echo behavior on interaction length $\xi$ and strength $\omega$ by examining the first moment of the echo, as done in Figure \ref{fig:first_moment_p_vs_xi}. The moment $\mu_1$ is given by
\begin{align}
    \label{eq:first_moment_def}
    \mu_1 = \sum_{t_{i}} (t_i - 2\tau)S(t_i)
\end{align}
for $S$ the signal magnitude, taken as the absolute value of the ensemble's planar magnetization.
In particular, for a perfectly symmetric echo $\mu_1 = 0$, so we can use this moment $(\mu_{s})$ as a measure of the effective asymmetry about $t = 2 \tau$.
When $\xi \approx 5$, most of the effects of very short range couplings have disappeared (only true when $\alpha_z = 0$), and the moments generally stay constant with increasing $\xi$.  The exception to this trend is the presence of the post echo phase locking, which we can see as a narrow band of near-zero first moment emerging in the long $\xi$ limit near $\omega = 3.3\,\Gamma$.

\subsubsection{Planar vs out-of-plane competition}
In Figure \ref{fig:echo_compare_all_xi_with_z}, we see that the introduction of even fairly weak $\omega_z$ relative to $\omega$ for small $\xi$ can result in strong attenuation of the echo.
Although a purely in-plane interaction also results in echo attenuation, in combination with the out-of-plane coupling the effect is much stronger and can result in almost total destruction of the echo.  
As $\xi$ increases, the echoes become more resilient to these dissipative effects: averaging over a larger area reduces the importance of local fluctuations.
This can be seen by comparing the $\xi = 6, 20$ echoes in Figure \ref{fig:echo_compare_all_xi_with_z}(a).
For a global interaction the local fluctuations disappear entirely and the presence of a small, non-zero $\omega_z$ has no effect.


\begin{figure}[!tbp]
    \centering
        \includegraphics[width=0.65\linewidth]{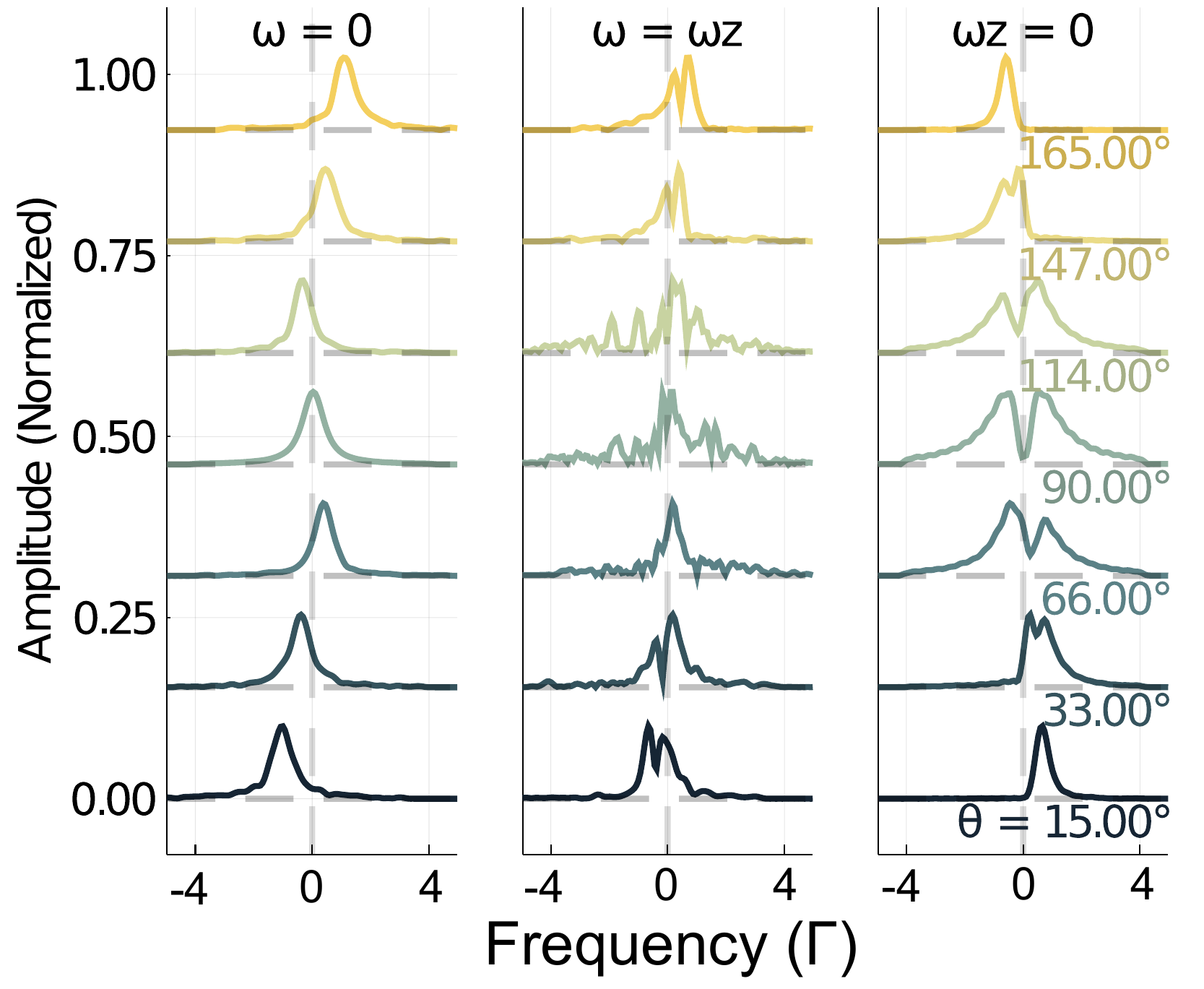}
    \caption{Variety of spectral behaviors observed in the spin echo for different flip angles with $\xi = 2.0$ and $(\omega, \omega_z)$ equal to $(0, 3.15\Gamma)$, $(3.15\Gamma, 3.15\Gamma)$, and $(3.15\Gamma, 0)$.}
    \label{fig:spec_vs_pulse_local_w_z}
\end{figure}

\subsubsection{Pulse angle dependence interaction induced dissipation/decoherence}
 We now discuss the effects of  spin  tip angles which differ from perfect pulses ($\theta \neq 90^\circ$), depicted  in Figure \ref{fig:spec_vs_pulse_local_w_z}. As demonstrated in the previous sub-section, the effects of moderate to long-range $\xi$ (6 and 20, respectively) are essentially identical to the global case. Therefore, to investigate effects of non-perfect pulses $(\theta \neq 90^\circ)$,     we focus on the short-range interactions ($\xi = 2$). Even in the short-range case, the observed spectral behaviors for $\omega_z = 0$ and $\omega = 0$ are essentially identical to the global case.  That is to say,  the short-range interaction only induces   a pulse dependent shift but does not alter the hole burning or phase locking effect. However,  including in-plane coupling and out-of-plane coupling together results in behavior similar to the global case, but with the addition of strong attenuation on account of the large values of $\omega_z$ needed to induce a noticeable shift. 
At these values of $\omega_z$, the highly local echoes are almost entirely attenuated and the observed spectral behavior is not a reliable indicator of the dynamics.  That is,  for $\omega_z = \omega$, the exact spectral shape as a function of  pulse angle   is dictated by the mean field interactions dragging the spins out of coherence.  This effect is present independently of a standard Linblandian dissipation term (due to rephrasing and/or relaxation) described in \mbox{Sec. \ref{Ldiss}}. The dissipation effects of the highly localized interaction are more dramatic as different regions of spins can experience dramatically different ``torques'' (to use a classical analog) from the interaction. 
Therefore, dissipation discussed in this section  does not arise from the Linbladian terms, but rather from the effect of the interaction itself. Nevertheless, this interaction dissipation  manifests as the effect that attenuates the spin echo and can thus be confused with standard sources of the loss of the spin coherence.  We note that this strong dissipation is analogous to the well known Pennington-Slichter (PS) result for the transverse dissipation rate $T_{2g}$ in the strongly correlated cuprates \cite{Pennington1991}. The PS result  calculates the expected decoherence time, $T_{2}$ relaxation time, due to electron-mediated nuclei-nuclei coupling, considering only a nearest-neighbor electron-nuclei interaction (e.g. ``short $\xi$'') and an experimentally derived electron spin susceptibility (MMP) in cuprate  superconductor YBa$_{2}$Cu$_{3}$O$_{7-\delta}$. While derivation of the  electron spin susceptibility   is in spirit equivalent to the mean-field approximation we make for electrons, numerical work is  not necessarily  identical. That is, our numerical work can be applied to longer range interactions, which are often encountered in correlated electron materials. 
Details of comparison of our simulation work to that of the PS are presented in \mbox{Sec. \ref{sec:ExpRel}}.

\subsubsection{Inclusion of dissipation}
\label{Ldiss}

One can also include explicit dissipation terms (separate from the effective attenuation caused by the interaction) into the Linbladian, either from dephasing or relaxation from thermal effects. 
If the dissipation   occur slower than $\tau$, we find it easily destroys the phase locking effects but leaves the hole burning effects nearly unaffected.
This can be understood by remembering that the phase locking depends on a relatively delicate balance of the spin states and interaction strength.
Dissipation causes the spins to break away from this locked state, not only reducing the number of phase locked spins but also reducing the strength of the interaction and its ability to keep those remaining spins locked, we expect to see the phase locking behavior become more rare and the hole burning effects of the echo to become more prominent.
In essence, additional dissipation will increase the point at which  the effect of long range interaction on the echo and spectral behavior approaches the global limit. That is,  longer range interactions effects will resemble those of   the very local interactions.

\subsection{Relevance to experiments in unconventional superconductors}
\label{sec:ExpRel}

\begin{figure}[!t]
    \centering
    \vspace{-0.5cm}
    \includegraphics[width=0.99\linewidth]{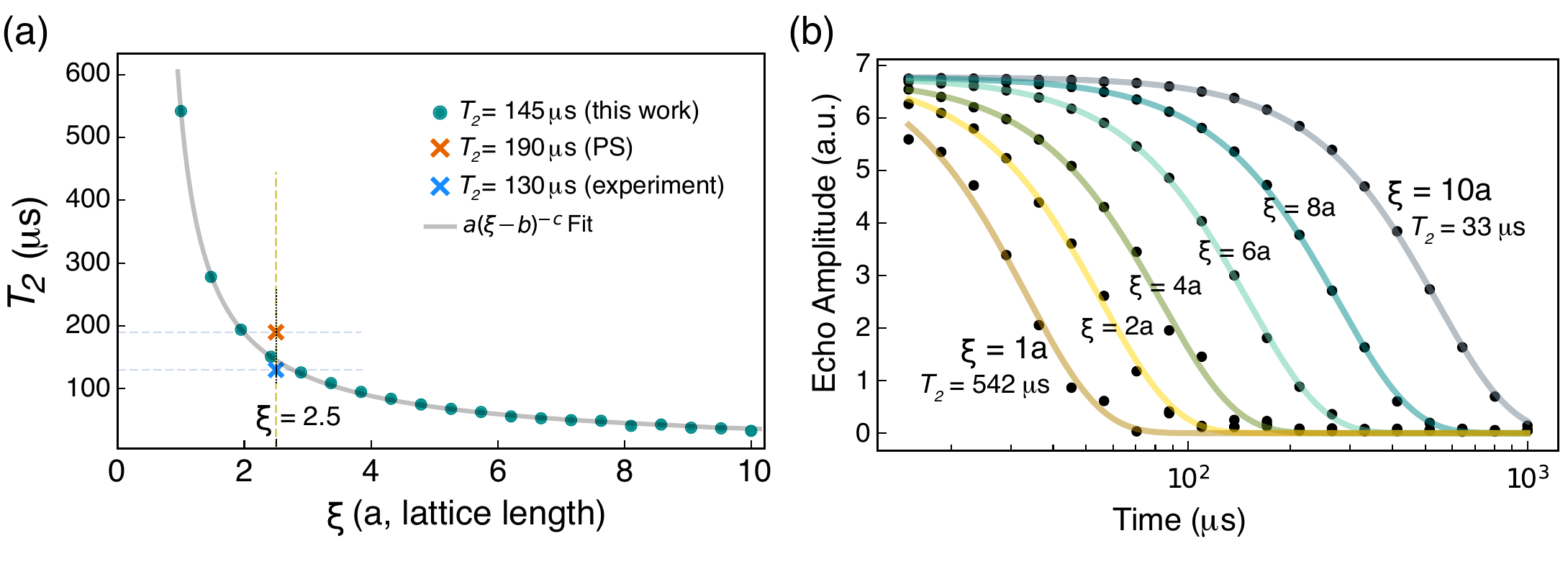}
     \vspace{-0.5cm}
    \caption{(a) The decoherence time $T_{2}$ induced by electron mediated nuclear spin-spin interaction as a function of electronic correlation length ($\psi$ given in the units of the lattice length $a$). Green circles were generated by our code for isotropic out-of-plane coupling with $\alpha_z= 0.003/\gamma$ and flip angle of $\theta = 75^{\circ}$. For this set of electron mediated coupling parameters, we find  
    $T_{2} \approx 145\, \rm{s}$ for $\xi=2.5\rm{a}$. 
    Experimentally measured value in cuprate  superconductor YBa$_{2}$Cu$_{3}$O$_{7-\delta}$ and that calculated in \mbox{Ref. \cite{Pennington1991}} for $\xi=2.5\rm{a}$ are denoted by $\times$. Dotted line is the error-bar for   calculated value in \mbox{Ref. \cite{Pennington1991}}.  (b) Time evolution/decay of spin echo amplitude for various values of  
    the electronic correlation length, $\xi$,  for the same electronic interaction parameters that mediate nuclear spin-spin coupling as in (a). The results were obtained for  40,000 spins ($200 \times 200$ lattice). 
    The spectral width of  100 kHz   centered at 10 MHz
gaussian kernel for the interaction was used. 
     Flip angle of 75 degrees was selected to accurately account for the  $z$ interaction. We note that  pulses are represented by   rotation operators  implying that  the pulses  a perfect pulse is perfect for all spins, not just those at the Larmor frequency. Lindblandian dissipation parameter  was set to zero to only account for dissipation from the interaction. 
 }
    \label{fig:PCT2}
\end{figure}

One of the main motivations for this numerical work was our aspiration to decide microscopic nature of Fulde-Ferell-Larkin-Ovchinnikov (FFLO) superconductivity.  
As alluded in the introductory paragraph,  
FFLO superconductivity \cite{PhysRev.135.A550, Matsuda_2007,Mitrovic08,Mitrovic2010,Mayaffre14} represent some of the most exotic correlated electron phases. In particular,  the most mysterious examples of the FFLO superconductivity is believed to be found in quasi 2D heavy fermion superconductor CeCoIn$_{5}$. The original experimental evidence for the emergence of the FFLO state below  temperature $T^{*}$, was the appearance   of a tiny NMR spectral 
signal observed at frequencies corresponding to the normal state signal. The results of the original NMR measurements reporting the composite lineshape closer to the normal state were interpreted as evidence for 
the emergence of a spatially modulated SC state \cite{PhysRevLett.97.117002,Mitrovic2008}. However, such composite lineshape and the relatively small shift changes between the low temperature   SC and normal state  were not presented in  any subsequent  NMR studies   \cite{PhysRevLett.97.117002,Mitrovic2008,Mitrovic08}. It is believed that  the composite lineshape observed in the vicinity of the $T^{*}$ is an artifact of erroneous/excesive RF excitation power
used in the original experiment  \cite{Mitrovic2008}. The  numerical results discussed in this section clearly indicate that this composite lineshape does not arise from the trivial effects of RF heating but rather signals dramatic change in anisotropy and correlation length  of electronic correlations as system transitions to the superconducting (``FFLO'') state. 
Specifically, as temperature is lowered through  $T^{*}$ and the system undergoes transition from the normal to the ``FFLO'', the electronic susceptibility becomes anisotropic (more 2D-like) and electronic  correlation length increases. Appearance of longer range anisotropic interaction induces hole-burning, \ie spectra with two apparent peaks as shown in \mbox{Figure \ref{fig:echo_compare_all_xi_with_z}}, and/or phase locking effects. 
This composite echo is evident in \mbox{Figure \ref{fig:spec_vs_pulse_local_w_z}} for pulse power exceeding $\theta \gtrsim 35^{\circ}$. 
As pulse power is lowered $(\theta \lesssim 20^{\circ})$, the composite lineshape transforms into a single peak echo that is shifted in frequency. At low temperatures, well below $T^{*}$ when superconducting/FFLO order parameter is fully developed  the putative FFLO phase is characterized by 2D susceptibility and larger correlation length. In addition, effective tip angle will decrease due to superconducting currents shielding effects. Therefore, deep in the FFLo state, for smaller tip-angle and in-plane interaction only, our results indicate that  one should observe pulse induced shift. Indeed, In the ``FFLO'' state of CeCoIn$_{5}$ enhancement of the shift beyond that in the low-field superconducting state was reported in \mbox{Ref. \cite{Mitrovic08}}. 
 
Next we discuss effects of interaction induced dissipation/decoherence. In \mbox{Fig. \ref{fig:PCT2}}, we plot calculated decoherence  time, that is generated by electron-mediated nuclei-nuclei coupling exclusively. 
Furthermore, we compare these to experimental findings in a cuprate  superconductor YBa$_{2}$Cu$_{3}$O$_{7-\delta}$.

 In the case of isotropic out-of-plane coupling, the dissipative effect of the interaction results in Gaussian relaxation of the spin echo, yielding an effective $T_{2}$ driven by the interaction rather than the typical mechanisms of nuclear spin dephasing. This effect can be seen clearly by simulating a set of echoes for varying echo time at a fixed interaction strength and coherence length, resulting in clear attenuation of the echo amplitude. This relaxation closely mirrors the relaxation calculated by Pennington-Slichter \cite{Pennington1991}. They  calculated the effective relaxation in YBa$_{2}$Cu$_{3}$O$_{7-\delta}$ driven by indirect coupling of the Cu nuclei mediated by electrons in the CuO$_{2}$ planes -- a mechanism broadly similar to the one modeled in this paper. In their case, they calculated the dependence of the relaxation time $T_2$ on the coherence length of electronic correlations that mediate the indirect coupling. As elaborated in  \cite{Pennington1991},  they found a $T_2$ of $190 \pm 75 \, \mu \rm{s}$   for a coherence length of $\xi = 2.5$ lattice lengths, with power-law $(T_{2} \sim  1/\xi)$ dependence of $T_2$ on the coherence length. Using our model, we can consider the dependence of the effective relaxation time driven by the out-of-plane coupling for varying $\xi$ and a fixed interaction strength to derive a nearly identical result, with the power $c=0.85$ and a $T_2$ of $145 \,  \mu \rm{s} $ at $\xi=2.5$ lattice lengths, well within their margin of error.  Additionally, because our model has a number of tunable parameters, it is likely that with the proper tuning our result would be in  complete agreement with experimental ones. 

Our findings here match well with our understanding of the interaction. The behavior of echoes under   interaction considered here  is driven broadly by the total interaction weight omega rather than the specific interaction strengths alpha. For a fixed interaction strength, a short coherence length reduces the impact of the interaction as it results in a reduced overall weight.  However, the increased effect of local fluctuations in the interaction driven by the short coherence length still results in relaxation. As the coherence length increases, the local fluctuations in the interaction decrease, but the overall strength of the interaction increases. We realize that high interaction strengths have a strong destructive effect on the spin echo regardless of coherence length; an effect we observe in this result, as the higher overall interaction weight more than offsets the reduced impact of local fluctuations. As evident in our result in \mbox{Fig. \ref{fig:PCT2}}, we would therefore expect to see a decrease in the relaxation time as the coherence length increases.

\section{Implementation and Program overview}
\label{sec:simulation}

Numerical simulations are a relatively common tool in the study of nuclear magnetic resonance.  There are several widely-used simulation packages available which are designed to be flexible tools for the simulation of NMR systems (see \cite{Candoli23} and references there in). 
Many of the simulations packages emphasize generality and accessibility, intending to present to the user the ability to simulate a wide range of NMR observables with minimal need for direct programming or familiarity with details of the package.
Though powerful   open-source software for simulation of observables in magnetic resonance experiments that can assist experimental research in the design of new strategies for the investigation of
fundamental quantum properties of materials, PULSEE \cite{Candoli23} allows the tuning of the parameters of several standard NMR terms rather than the introduction of general interactions between the spins. 
 When it comes to the simulation of NMR observables in strongly correlated phases of matter, especially if one is interested in exploring the effect of long-range interactions, no existing packages provide a good starting point.

Full many-body simulations of spin-$1/2$ nuclei in a solid state system are impractical for large numbers of nuclei $N$.
This is because the Hilbert space (basis) for such a problem consists of all possible arrangements of the $N$ spins, e.g. $\{s_1, s_2, \dots , s_N\}$ where $s_i = \{\uparrow, \downarrow\}$.
Exact-diagonalization of the many-body Hamiltonian thus has the famously intractable $2^N$ scaling in the matrix size.
This exceeds current computational capability around $N=20$. 
In the case of a mean field model, which treats each spin as its own basis element instead of the $N$-spin configuration as a basis element, the matrix size goes from $2^N$ to $2N$.
This linear scaling makes the problem feasible to store in memory, but is still computationally challenging to diagonalize or time-evolve as $N$ approaches hundreds of thousands of spins.

One of the primary goals of this work is the development of a software package that can accurately and efficiently study large ensembles of spins with novel interactions.  Our program focuses on the simulation of large ensembles of spins whose interactions can be modelled on the mean field level.  The particular form of the interaction is not essential to the broad structure of the simulation; however, the interaction is hard coded to optimize performance, and changing the interaction on a structural level requires directly changing of the simulation code.  We additionally take advantage of GPU computing-- a tool not widely implemented yet in time-domain NMR simulations-- to accelerate the program, allowing for the efficient simulation of large ensembles of spins.

In the process of developing the simulation package and studying the model, we wrote three versions of a spin echo simulation \cite{spin_sim_repo}.  The first iteration, written entirely in Julia and utilizing no GPU programming (referred to moving forward as simply \textit{the Julia simulation}), was primarily used in our initial exploratory investigations of the model; it is essentially a direct translation of the model described above, and it primarily serves as a tool to quantify the performance of more optimized programs \cite{bezanson2017julia}.  We use the other two simulations for our numerical experiments, one written in Julia with some GPU functionality (which we call the \textit{GPU-Julia simulation}) and one written in CUDA (the \textit{CUDA simulation}) \cite{besard2018juliagpu, besard2019prototyping, cuda}.  In the following sections, we discuss these two GPU simulation packages in depth, focusing on their key calculation steps, accuracy, and performance.

In the following sections, we review the basic structure of the Julia and CUDA simulations, some of the more unique elements of their implementation, and their performance.  We review the two non-trivial calculations involved in the simulations, namely the calculation of the local magnetization $M_{d,L}^{(i)}$ and the calculation of the matrix exponential in the CUDA implementation.  We then examine the performance of the three simulations (Julia, GPU-Julia, and CUDA) and examine the accuracy of the handwritten algorithms in the CUDA simulation by comparing its results to those from the GPU-Julia simulation.

\subsection{Code Structure}
\label{sec:code_implementatin}

Both simulations implement the same model and calculate the output of a basic spin-echo experiment.  The calculated signal and output of the simulations is the net magnetization along each axis, $M_d$, given by
\begin{align}
    M_d = \sum_{i = 0}^{N}\langle I_d^{(i)} \rangle_R \quad \textrm{for $i = z, +$}
\end{align}
where we save the value of the ``rotating'' expectation value, as defined in Section \ref{sec:rot_frame}.

To perform the simulation, we first generate and initialize an ensemble of density matrices as well as a number of variables that are held constant throughout the experiment.
Then, we apply the spin echo protocol: (1) flip the spins an angle $\theta$ (\ie a rotation of each density matrix by an angle $\theta$), (2) time evolve under our model for a time $\tau$, (3) apply rotation to flip the spins by an angle $\varphi$, (4) and then time evolve for a time $2\tau$.
The values of $M_z, M_+$ at each time step and after each pulse are stored for analysis.

Although in principle every variable in the system can be adjusted freely, the simulations are designed to accommodate easy adjustment of the most commonly changing variables: the interaction strengths $\alpha$, the length scaling function $f$, the dissipation coefficients $\Gamma_d$, the lattice dimensions $n_x$, $n_y$, the echo time $\tau$ and the time step $dt$, and the pulse angles $\theta$ and $\varphi$. 
These ``free'' parameters for a given set of trials are loaded from a parameter file (a simple text file) at the start of an experiment. In the CUDA simulation, the parameters are loaded from a pair of parameter files: one denoted ``simulation'' parameters and one denoted ``echo'' parameters.  Simulation parameters are those upon which memory allocation depends, such as $n_x$ or $\tau$, and echo parameters are those upon which it does not.  In the CUDA simulation, the loop over trials is performed with a nested loop, where the outer loop iterates over the simulation parameters and the inner loop over echo parameters.  This structure allows us to perform the memory allocation once per set of echo parameters by reusing and overwriting memory from completed trials, minimizing the time spent on memory management.

In the GPU-Julia simulation, we have limited control over how the calculation is parallelized on the GPU.  However, Julia's CUDA package has highly-optimized parallelization schemes for many basic matrix operations built-in. 
In CUDA, we have direct and complete control over how a calculation is divided up across the threads on the GPU.  As our model consists of thousands of independent manipulations of small matrices (typically two or four dimensional), we parallelize the system by particle: we assign each particle to a single thread which performs the full calculation for said particle for a single time step.

\subsection{Local Magnetization: $M_{d,L}^{(ij)}$}
\label{sec:local_mag_calc}

The calculation of the local magnetization uses an $n_x \times n_y$ matrix $S$ known as a ``stencil'' (inspired by finite element methods), where $n_x$ and $n_y$ are the lengths of the lattice along the $x$ and $y$ axis, respectively. Here, we will index particles based on their $x$ and $y$ coordinates:  $M_{d,L}^{(i)}$ becomes $M_{d,L}^{(ij)}$ and our previous notation $f(r_{ij})$ becomes
\begin{gather}
    f(r_{ij}) \to f(r_{ij,k\ell})
\end{gather}
where $ij$, $k\ell$ refer to each particle's $x$ and $y$ coordinates.  The stencil $S$ is defined by
\begin{gather}
    S_{ij} \equiv f(r_{00, ij})
\end{gather}
with indexing now starting at zero.  Additionally, $S_{00} \equiv 0$, a condition which prevents self-coupling.  The stencil $S$ is, by default, ``centered'' at $(0,0)$.  By taking advantage of our periodic boundary conditions, we can shift the stencil to be centered at any given point $(k,\ell)$.  Letting $S^{(k\ell)}$ designate a stencil centered at $(k,\ell)$, we can implement the following transformation between $S$ and $S^{(k\ell)}$:
\begin{equation}
    \begin{gathered}
    S^{(k\ell)}_{ij} = S_{i'j'} \\
    i' = (i - k)\,\textrm{mod}\,n_x \\
    j' = (j-\ell)\,\textrm{mod}\,n_y \, .
    \end{gathered}
\end{equation}
Additionally, we use an $n_x \times n_y$ matrix $M^{eval}_d$ which contains the expectation value of $I_d$ of each particle on the lattice.  We define $M^{eval}_d$ by
\begin{gather}
    \left( M^{eval}_d \right)_{ij} \equiv \langle I_d^{(ij)} \rangle_R \, .
\end{gather}
With these definitions in place, we can see that the definition of $M_{d,L}^{(ij)}$ as given in Equation \ref{eq:local_m_def} is equivalent to the sum of the element-wise multiplication of $M_d^{eval}$ and $S^{(ij)}$.

The fundamental process used to compute $M_{d,L}^{(ij)}$ is the same across all three versions of the simulation (Julia, GPU-Julia, and CUDA), although the specifics of the implementation vary.  In all three cases, we calculate the value of $S$ a single time at the start of a trial, as the calculation represents one of the slower processes in the simulation.  In the Julia and GPU-Julia simulations, we also calculate the full set of shifted stencils $S^{(ij)}$ for all particles prior to performing the time evolution.  The set of shifted stencils is saved to a four dimensional array $\hat{S}$, defined by
\begin{align}
    \hat{S}_{ijk\ell} = S^{(ij)}_{k\ell} \, .
\end{align}

In the Julia simulation, the process is implemented directly with a simple loop over the particles, with each iteration performing the element-wise multiplication and sum between $S^{(ij)}$ and $M_d^{eval}$.  This process is done solely on the CPU and is by far the slowest step of the simulation, accounting for over $90\%$ of the computation time of the Julia simulation.

In the GPU-Julia simulation, we perform the calculation of $M_{d,L}^{(ij)}$ on the GPU.
By restructuring $M_d^{eval}$, we can perform the calculation of $M_{d,L}^{(ij)}$ for all particles with two matrix operations without any need for loops.  Each of these operations will be automatically optimally parallelized by the CUDA library's default parallelization structures.  However, accessing these optimized functions comes at the cost of dedicating a non-trivial amount of computation time to re-structuring the calculation to make it compatible with these functions.  Additionally, there is the further loss of efficiency due to the need to move memory back and forth between the GPU and CPU before and after the calculation of $M_{d,L}^{(ij)}$.

In order to re-structure the calculation, we first create a four dimensional array $\hat{M}_{d}^{eval}$ where each slice is a copy of $M_d^{eval}$:
\begin{gather}
    \big(\hat{M}_{d}^{eval}\big)_{ijk\ell} \equiv \big(M^{eval}_{d}\big)_{k\ell} \, .
\end{gather}
The element-wise multiplication step for all particles can then be performed with a single element-wise multiplication between $\hat{M}_d^{eval}$ and $\hat{S}$.  Performing the sum is equally simple, using \textit{mapreduce} on the two dimensional slices of the resulting four dimensional array.  Julia's CUDA library automatically uses routines from CUDA's cuBLAS libraries for these operations, routines which are designed to take advantage of the GPU's potential for large-scale parallelization and can be assumed to be highly optimized.  Despite the additional need for memory management and the creation of $\hat{M}_d^{eval}$, for a $100 \times 100$ lattice this restructuring results in a nearly ten-fold reduction in computation time, including the time spent on the additional steps.

The implementation of the $M_{d,L}^{(ij)}$ calculation in the CUDA simulation is the most direct of the three approaches, as the abilities to customize the parallelization structure and directly manage memory give us more freedom with how to structure the calculation.  When calculating $M_{d,L}^{(ij)}$, each particle is given an index (which is also its thread index).  CUDA does not facilitate multidimensional arrays well, so all arrays are flattened in row-major order.  With this in mind, we can convert the spin/thread index, which we will call $q$, into a pair of indices which point to the particles position on the lattice:
\begin{equation}
    \begin{gathered}
    j = q\,\textrm{mod}\,n \\
    i = \frac{q - j}{n}
    \end{gathered}
\end{equation}
where we are operating on a square lattice with $n_x = n_y \equiv n$.  We then define two new sets of indices: $k,\ell$ and $k',\ell'$.  These indices are related by
\begin{equation}
    \begin{gathered}
    k' = (k + i) \, \textrm{mod} \, n \\
    \ell' = (\ell + j) \, \textrm{mod} \, n \, .
    \end{gathered}
\end{equation}
In this relation, $i,j$ refer to particle position, $k,\ell$ refer to the stencil, and $k',\ell'$ refer to $M_d^{eval}$.  Adding $i,j$ to $k,\ell$, modulo the lattice size, applies the shift previously calculated in $\hat{S}$.  We calculate $M_{d,L}^{(ij)}$ by summing over $k,\ell$:
\begin{gather}
    M_{d,L}^{(ij)} = \sum_{k = 0}^{n-1} \sum_{\ell = 0}^{n-1} S_{k\ell} \big( M_d^{eval} \big)_{k'\ell'} \, .
\end{gather}
The ability to customize the parallelization structure allows this version of the calculation to bypass the need to calculate $\hat{M}_d^{eval}$ or $\hat{S}$ as in the GPU-Julia version while still allowing us to parallelize over the indices $i$ and $j$.  The CUDA simulation's calculation improves upon the performance of the GPU-Julia calculation by an additional twenty times for a $100 \times 100$ lattice.

\subsection{Matrix Exponent}
\label{sec:mat_exp}

Although Julia has a built in matrix exponent, CUDA lacks such a function.  To calculate the matrix exponent in the CUDA simulation, we use a handwritten version of the scaling-and-squaring algorithm, the same algorithm used by both Julia and MATLAB \cite{bezanson2017julia, 10.1137/04061101X}.  The details of this algorithm are beyond the scope of our discussion, but we will briefly outline its steps to give context to the details of our implementation.

The scaling-and-squaring algorithm uses a Pad\'{e} approximant for the matrix exponent $e^A$.  In order to improve accuracy, the algorithm first reduces the norm of the matrix $A$ by scaling it by a factor $2^s$, with the choice of $s$ dependent on the initial norm of the matrix $A$.  The numerator $P_m$ and denominator $Q_m$ of the Pad\'{e} approximant are then calculated (each a polynomial function of the rescaled matrix of order $m$), and a linear solver is used to solve for the approximant itself.  The result is then squared $s$ times to reverse the initial scaling, resulting in an approximation of $e^A$.

Although CUDA has accurate and efficient linear solvers via the cuSOLVER library, the solvers in this library are optimized around parallelizing over a single large matrix operation rather than an ensemble of small matrix problems, much like the cuBLAS libraries \cite{cuda}.  In order to parallelize over the ensemble, we use a handwritten Gaussian elimination and backwards substitution algorithm for our linear solver.  Although such an algorithm scales extremely poorly with matrix size, our simulation is designed for the modelling of independent, single-particle density matrices, which are typically two or four dimensional.  The poor scaling is therefore not an issue in this context.

In order to examine the accuracy of our matrix exponential algorithm, we compared it to Julia's built in algorithm for matrices with elements from multiple orders of magnitude.  We focused on $4 \times 4$ matrices for the analysis, as the majority of our examination of this model occurs for spin $\nicefrac{1}{2}$ systems (and the simulation is performed in Liouville space).  To generate a matrix $M$, we first generate a set of four complex eigenvalues and save them in a diagonal matrix $D$.  We then conjugate this matrix by a unitary matrix $U$, generated via exponentiation of a random Hermitian matrix $H$:
\begin{equation}
    \begin{gathered}
        \label{eq:matrix_creation_def}
        D = \textrm{diag}\,(\lambda_1, \dots, \lambda_4) \\
        U = e^{iH} \\
        M = UDU^\dag \, .
    \end{gathered}
\end{equation}
This process gives us a random complex matrix $M$ with a known set of eigenvalues.  We exponentiate this matrix using our CUDA algorithm and compare the result to the same matrix exponentiated using Julia's built in function.  We analyze matrices whose eigenvalues fall in the ranges $10^{-p}, 10^{-p+1}$, for $p$ ranging from 9 to 2.  We limit ourselves to eigenvalues whose magnitude is strictly less than 1 due to the reliance of both the Julia algorithm and the CUDA algorithm on high order products of the matrix being exponentiated, resulting in poor behavior of both algorithms for matrices whose elements are larger than 1.  We additionally analyze a set of matrices with mixed value eigenvalues, drawn at random from the full range $10^{-9}$ to $10^{-1}$.

The error of the exponentiation of a given matrix using the CUDA algorithm, $U^{(c)} = e^{M}$, relative to the same calculated using the Julia algorithm, $U^{(j)}$ was calculated using the relative error of the real and imaginary parts separately.  In particular, the error of a given matrix element $(U^{(c)})_{ij}$ has two values, one for the real part and one for the imaginary part, given by
\begin{equation}
    \begin{gathered}
    \label{eq:matrix_error_def}
    \delta_{r,ij} = \frac{\textrm{Re}\,(U^{(c)})_{ij} - \textrm{Re}\,(U^{(j)})_{ij} }{\textrm{Re}\,(U^{(j)})_{ij} }
    \end{gathered}
\end{equation}
and the equivalent $\delta_{i,ij}$ for the imaginary part.  The ``overall error'' for a given matrix is given by considering both the average across all matrix elements of the real and imaginary parts as well as by considering the average \textit{magnitude} of the error of the real and imaginary parts across all matrix elements.

There were no identifiable trends in the error relative to the magnitude of the largest eigenvalue of the matrix.
Across all regimes studied, the error was consistently small.
The real part had an average error on the order of $10^{-6}$ and the imaginary part on the order $10^{-7}$.
The distribution of error about these averages was reasonably tight, with the a standard deviation in the error on the order of $10^{-7} and 10^{-8}$,  respectively. 
The source of this small error is likely from differences in the choice of the Pad\'e approximant used in the scaling-and-squaring algorithm, or different levels of precision in the underlying matrix routines (which we discuss below).
Nonetheless, these small errors seem to not contribute in any noticeable way to the simulated spin echoes.
Thus, we can take our handwritten matrix exponential algorithm in CUDA to be equivalent to the Julia implementation for the purposes of our spin simulations.

Although most calculations across both versions of the simulation are performed in single precision (as GPU performance is greatly improved by using single precision), the calculation of the matrix exponent in the Julia version of the simulation is performed in double precision.  Although this somewhat reduces the performance of the Julia simulation, it improves accuracy-- when applied to matrices stored in single precision with small matrix elements (e.g. eigenvalues on the order of $10^{-9}$ to $10^{-8}$). We found that Julia's matrix exponential shows significant divergence relative to the CUDA algorithm, MATLAB's matrix exponential algorithm, and even from the same Julia function calculated in double precision.  In order to avoid errors from cases like these, we perform the matrix exponentiation in double precision in Julia, converting back to single precision for the other calculations in the simulation.

\subsection{Performance}

To profile our spin echo simulation software, we run fifty trials for each implementation over a range of values of maximum strength of interaction $(\Omega)$, as defined in Equation \ref{eq:big_omega_def}, in order to account for potential fluctuations in performance for different simulation parameters.  Each trial was run on a $100 \times 100$ lattice, the standard lattice size we use for our simulations, with a sampling rate of $\nicefrac{1}{dt} \approx 3$MHz.  We first examine the performance of the Julia simulation to establish a baseline against which to compare the improvements gained in the GPU-Julia and CUDA simulations.

\begin{table}
    \centering
    \begin{tabular}{c|c|c}
    	Calculation/Process	    &	\begin{tabular}{@{}c@{}} Time per \\ Execution (ms) \end{tabular}	&	$\%$ of Total Time	\\	\hline
    	Calc. Constants		    &	$	381.161	$	&	$	0.177	$	\\	
        $	M_d^{eval}	    $	&	$	0.158	$	&	$	0.018	$	\\	
        $	M_d	            $	&	$	0.070	$	&	$	0.008	$	\\	
        $	M_{d,L}^{(ij)}	$	&	$	722.676	$	&	$	81.421	$	\\	
        $	H	            $	&	$	30.802	$	&	$	3.456	$	\\	
        $	U	            $	&	$	131.007	$	&	$	14.698	$	\\	
    	Propagate		        &	$	1.987	$	&	$	0.223	$	\\	\hline
    	Total Time		        &	$	419825	$	&	$	-	$	\\	

    \end{tabular}
    \caption{Profiling of the major steps of the Julia simulation.  $\%$ of total time is calculated by multiplying the time per execution by the number of executions (typically either $1$ or $3n_\tau$), divided by the total time for the trial.}
    \label{tab:julia_cpu}
\end{table}

The performance of the Julia simulation is summarized in Table \ref{tab:julia_cpu}.  The ``Calculation of Constants'' step summarizes the total time spent on variables that remain constant throughout the trial and are typically only calculated once or twice, primarily the stencils $S$ and $\hat{S}$.  These steps can be some of the slowest on a per-execution basis, but they are performed a fraction of the times as the other steps.

On average, trials took almost 420 sec each.  By far the most dominant contribution to this poor performance is the calculation of $M_{d,L}^{(ij)}$ for $d = z, +$.  The poor performance of this step is unsurprising, given that it essentially must iterate over $100,000,000$ lesser calculations.  The performance of this step is only better than a serial calculation of 100,000,000 multiplications and 100,000,000 additions (on complex floats) by a factor of around $10$. This was tested locally, where a single multiplication or addition of two complex floats takes around 25 ns, whereas performing the $M_{d,L}^{(ij)}$ calculation for a single particle takes around 50 $\mu$s.

\begin{table}
    \centering
    \begin{tabular}{c|c|c}
    	Calculation/Process		&	\begin{tabular}{@{}c@{}} Time per \\ Execution (ms) \end{tabular}	& $\%$ of Total Time \\	\hline
    	Memory Management		&		16.340		&		7.812		\\	
    	Calc. Constants		    &	$	340.524	$	&	$	0.671	$	\\	
        $	M_d^{eval}	    $	&	$	0.245	$	&	$	0.118	$	\\	
        $	M_d	            $	&	$	0.124	$	&	$	0.059	$	\\	
        $	M_{d,L}^{(ij)}	$	&	$	9.481	$	&	$	4.551	$	\\	
        $	H	            $	&	$	35.578	$	&	$	17.008	$	\\	
        $	U	            $	&	$	141.359	$	&	$	67.575	$	\\	
    	Propagate		        &	$	4.615	$	&	$	2.206	$	\\	\hline
    	Total Time		        &	$	98528	$	&	$	-	$	\\	

    \end{tabular}
    \caption{Profiling of the major steps of the GPU-Julia simulation.  $\%$ of total time is calculated by multiplying the time per execution by the number of executions (typically either $1$ or $3n_\tau$), divided by the total time for the trial.}
    \label{tab:julia_gpu}
\end{table}

The GPU-Julia simulation is essentially the same as the Julia simulation, with the exception of the calculation of $M_{d,L}^{(ij)}$, a fact we see reflected in the similar per-execution performance of many of the GPU-Julia simulation's steps, contained in Table \ref{tab:julia_gpu}.  By moving the calculation of $M_{d,L}^{(ij)}$ to the GPU, however, we see a significant boost in performance.  The calculation of $M_{d,L}^{(ij)}$, including the additional time spent calculating $\hat{M}_d^{eva}$, shows an improvement of almost $80$ times.  We suffer the additional loss of needing to manage memory between the GPU and the CPU, but this loss (an additional $16$ ms per calculation of $M_{d,L}^{(ij)}$) is more than made up for by the improvement in the calculation itself.  Combined with other minor gains in some of the stencil calculations (accomplished by performing them on the GPU), reduce the per-trial time from 420 sec to $98.5$ s, about a $4\times$ speedup.

However, given the majority of the remaining computation time in the GPU-Julia simulation is in the calculation of the propagator, there is little room for further improvement in performance.  As mentioned previously, CUDA does not have a built in matrix exponential function so any additional improvements to the GPU-Julia simulation necessitate some amount of direct CUDA programming, at which point it is more optimal to translate the entire simulation into CUDA to 
 eliminate some of the overhead accrued by using hybrid GPU-CPU programming.

\begin{table}
    \centering
    \begin{tabular}{c|c|c}
    	Calculation/Process		& \begin{tabular}{@{}c@{}} Time per \\ Execution (ms) \end{tabular}	& $\%$ of Total Time \\	\hline
    	Memory Management		&	$	193.216	$	&	$	4.674	$	\\	
    	Calc. Constants		    &	$	2.788	$	&	$	0.067	$	\\	
        $	M_d^{eval}	$	    &	$	0.008	$	&	$	0.088	$	\\	
        $	M_d	$	            &	$	1.126	$	&	$	12.882	$	\\	
        $	M_{d,L}^{(ij)}	$	&	$	3.543	$	&	$	40.372	$	\\	
        $	H	$	            &	$	0.048	$	&	$	0.548	$	\\	
        $	U	$	            &	$	3.594	$	&	$	40.943	$	\\	
    	Propagate		        &	$	0.037	$	&	$	0.426	$	\\	\hline
    	Total Time		        &	$	4134	$	&	$	-	    $	\\	
    \end{tabular}
    \caption{Profiling of the major steps of the CUDA simulation.  $\%$ of total time is calculated by multiplying the time per execution by the number of executions (typically either $1$ or $3n_\tau$), divided by the total time for the trial.}
    \label{tab:cuda_profile}
\end{table}

The CUDA simulation shows marked improvements across the board, with an overall reduction in per-trial computation time of almost 25 times relative to the GPU-Julia simulation, from over $98$ sec to just over $4$ sec per trial.  Memory management is by far the most costly per-execution step, but given memory management occurs on average less than once per trial (since it is performed in the outer loop, a large fraction of the trails re-use the same set of allocated memory), it has little impact on the overall computation time.  Parity has been achieved between the two most costly calculations, $U$ and $M_{d,L}^{(ij)}$, with each consuming around 40\% of the computation time.  Performance decreased in the calculation of the signal, $M_d$, as this calculation is inherently serial and by nature performs poorly on the GPU.  However, the calculation of $M_d$ consumes only $12\%$ of the total computation time, and methods such as \textit{atomicAdd} exist that could be used to improve its performance, with some additional development.

Overall, performance improved from $420$ sec per trial to $4$ sec per trial in the transition from Julia to CUDA, which is more than a $100\times$ speedup.  The CUDA simulation can be optimized further, as we use several inefficient algorithms that could be improved with 
 further development.  Better parallelization schemes could be incorporated, as our scheme uses at most 10,000 threads per calculation.  However, there are limits to the improvements available from increased parallelization.  Local memory requirements, based on the number of variables declared on each thread during the calculations and the number of threads in a thread block, limit the number of threads that can be launched on a single calculation.  Furthermore, the total number of threads on the GPU set an absolute upper bound on parallelization.  For example, a reasonably priced commercially available GPU, such as the GTX 1080 Ti, on which much of the development of this simulation was performed, has \textit{at most} 57,344 threads available \cite{forum_post}.  Further gains on a single GPU are limited, and higher level parallelization schemes likely merit cross-GPU parallelization, despite the additional overhead of managing memory between multiple GPUs.

\subsection{Validation}
\label{sec:validate}

In order to validate the accuracy of the handwritten algorithms in the CUDA simulation as a whole, and to make sure the small variations between the two simulations do not affect results, we examine the error between the spin echoes generated by the CUDA simulation to those generated by the GPU-Julia simulation.  Although the individual algorithms in the two simulations have only small variations between them, we do a comparison of the echo behavior to guarantee that these small errors do not propagate throughout a trial and accumulate to generate a noticeable difference.  Furthermore, an examination of the echoes generated by each simulation can speak to the numerical stability of the model and serve as an indication of the degree to which the results of the model are independent from numerical fluctuations in the parameters and algorithms.

In order to analyze the similarity of the results of each simulation, we simulate a set of echoes with $\omega$ ranging from $\Gamma$ up to $4.5\, \Gamma$.  We use the standard $\Gamma$ scaling for our other parameters: $\tau = \nicefrac{5}{2\Gamma}$, $\Gamma \, dt = \nicefrac{1}{10}$, as well as an ensemble with $n = 10,000$ spins.  We use a fixed frequency sampling for all trials, rather than resampling the frequency spectrum for each echo, to eliminate the variations between the echoes with the same parameters but different random sampling.

\begin{figure}[t]
    \centering
    \includegraphics[width=0.75\linewidth]{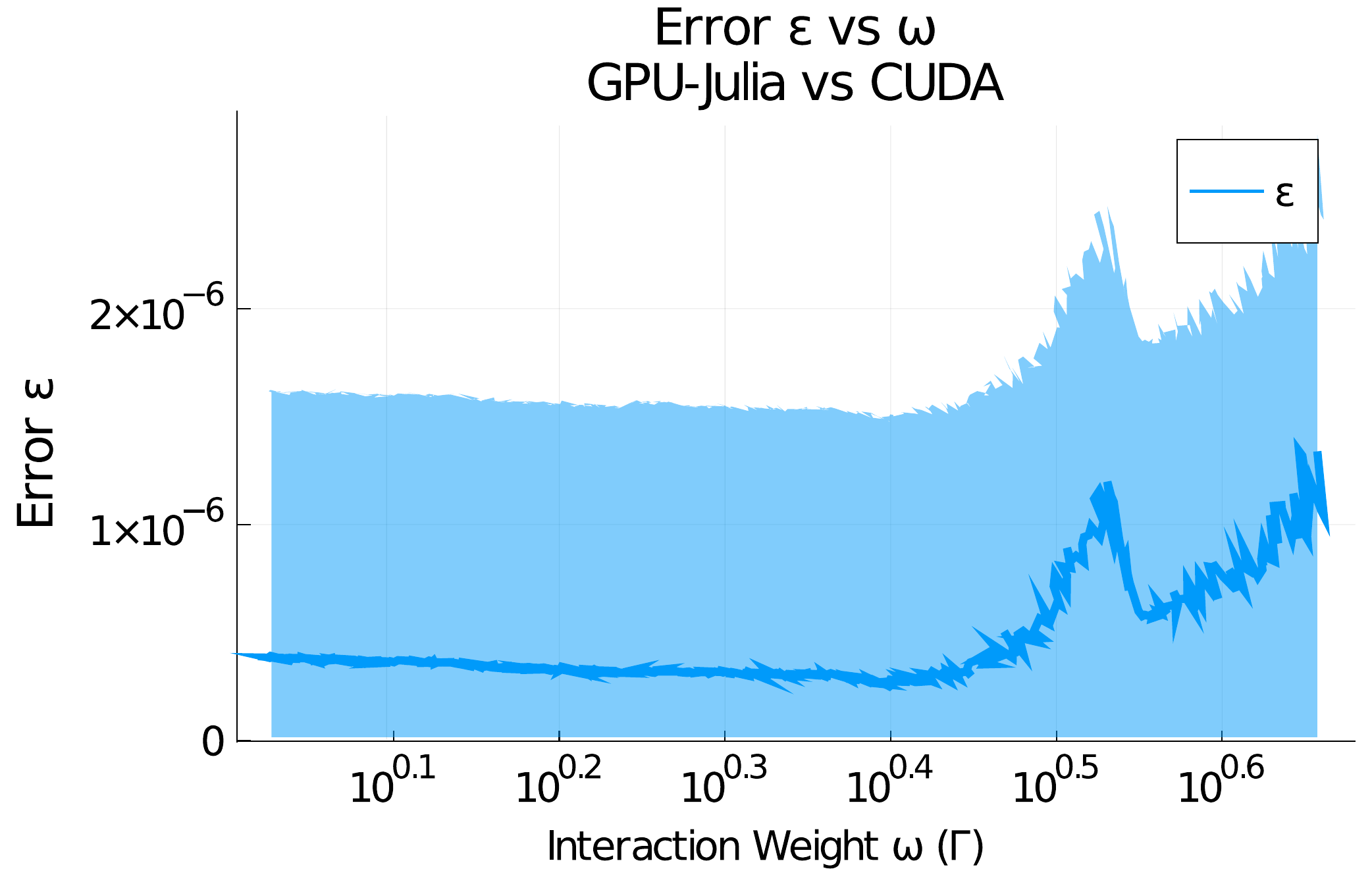}
    \caption{Error $\varepsilon$ between echoes calculated by the CUDA simulation and the GPU-Julia simulation, plotted versus $\omega$ as a fraction of $\Gamma$.  The shaded region represents one standard deviation about the error.}
    \label{fig:valid_error}
\end{figure}

We calculate the error $\varepsilon$ between the signal from an echo calculated using the CUDA simulation, $S^c(t)$, and the signal from an echo calculated using the GPU-Julia simulation, $S^j(t)$ by considering the magnitude of the error at each point, averaged over the whole echo.
\begin{align}
    \label{eq:valid_error_def}
    \varepsilon = \frac{1}{L}\sum_{i = 1}^{L}\left| S^c(t_i) - S^j(t_i) \right|
\end{align}
for $L$ the length of the echoes in question.  The value of $\varepsilon$, as well as its standard deviation, calculated using the individual errors of each point of each echo, is shown in Figure \ref{fig:valid_error}.

It is immediately apparent that the differences between the two simulations are negligible.  The average error magnitude is on the order  of $10^{-7}$ and the spread in error is consistently on the order of  $10^{-6}$.  Even at multiple standard deviations from the mean, this represents an error of a fraction of a fraction of a percent, on the order of $0.001\%$.  For all intents and purposes, the simulations can be considered identical.  We do see an increase in the error, although it remains on the order of $10^{-6}$, as $\omega$ increases towards values near $\Gamma$. 
This region is more sensitive to the exact values in the simulation, as the accuracy of the approximations used in the simulation becomes more important at higher values of $\omega$. 
This uptick in the error is therefore expected given small differences in the numerical details of the simulations, such as precision types and the solvers used in the matrix exponential algorithm. 
We recall the accuracy of our matrix exponential algorithm is on the order of $10^{-8}$ (Section \ref{sec:mat_exp}) which is consistent with the error seen between the echoes of the two simulations when accumulated over $2 \tau = 50 dt$ time steps.

\subsection{Baseline Numerical Error}
\label{sec:baseline_error}

Before we turn to the convergence of the model, we address one additional source of error that arises from the technical details of the simulation itself.  Specifically, the CUDA simulation uses floating point precision throughout to maximize performance.  However, this practice  introduces the possibility for errors on account of the cutoff and rounding of single precision.  In order to minimize these errors,  we employ custom arithmetic operations built out of CUDA's intrinsic functions which provides direct control over the rounding of the floating point values.  

However, in one specific situation, there is still an observable error that arises from a combination of floating point rounding and our data transfer protocols.  In the case of extremely high sampling rate, that is very small $dt$, the vast number of matrix exponential products can result in an accumulated error which is small but observable.  This primarily arises when comparing echoes with different but equivalently small time steps, for example an echo with sampling rate $\Gamma \, dt = 1 \times 10^{-5}$ and one with sampling rate \mbox{$\Gamma \, dt = 2 \times 10^{-5}$}.  In this situation -- which would almost never be relevant to an experimental application of the simulation, given the excessive computation time and diminishing returns in terms of improving accuracy -- there is a ``baseline error'' in the accuracy of the echo shape.

\begin{figure}[!tbp]
    \centering
    \includegraphics[width=0.75\linewidth]{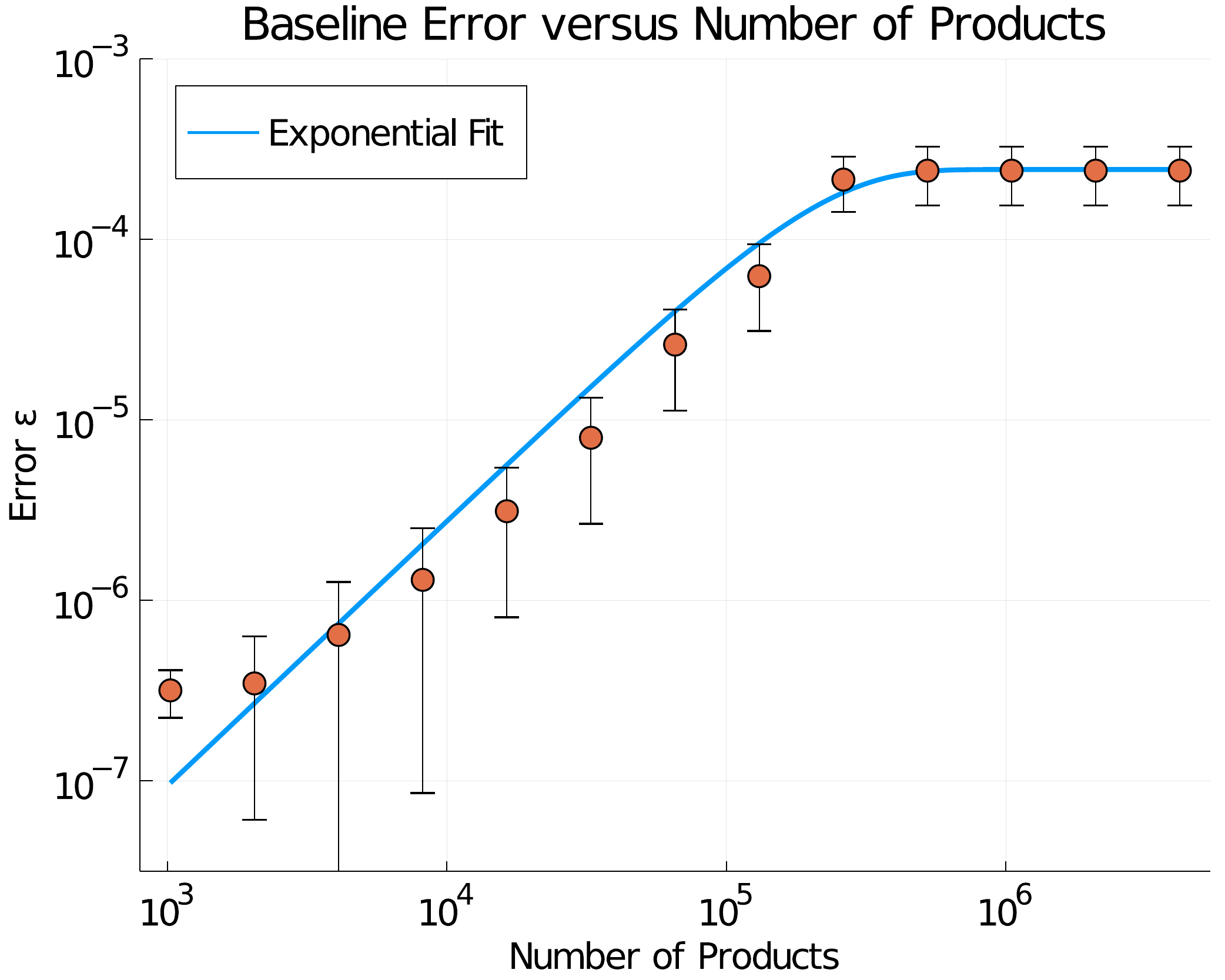}
    \caption[Baseline numerical error]{Baseline error $\varepsilon$ versus $n$.  The data is fit with an exponential fit $\varepsilon \propto 1 - e^{-n/c}$ for a constant $c$.  The error bars give the $45\%$ standard deviation of the error $\varepsilon$ about its mean over the 100 samples.  The error bars are scaled down to $45\%$ their value on account of low $n$ errors having a spread on the same order as their error and thus being unable to be plotted on a log scale.}
    \label{fig:base_error}
\end{figure}

In order to quantify this error, we use the matrix exponential function implemented in the simulation to compute a certain matrix in several ways.  Specifically, we perform the following computation
\begin{align}
    \label{eq:baseline_error_calc}
    e^{A} = \prod_{i = 1}^{n}e^{\nicefrac{A}{n}}
\end{align}
for $\log_2 n = 10$ up to $22$ to represent the range of $dt$ we study in Section \ref{sec:dt_conv}.  For our discussion, let $E_n$ be the calculation of $e^A$ using $n$ products.  We perform this calculation for 100 different matrices $A$ all of which are complex valued and randomly generated with elements representative of the order of magnitude we see in our simulations.  Additionally, the structure of $A$ is the same as the Hamiltonian in our system.  We then calculate the error $\varepsilon$ via
\begin{align}
    \label{eq:base_error_def}
    \varepsilon(n) = \frac{1}{d^2}\bigg|\sum_{i,j}^d \left( \left( E_n \right)_{ij} - \left(E_1\right)_{ij}\right) \bigg|
\end{align}
for $d$ the dimension of the matrices, here equaling to $4$.  We consider the average value of $\varepsilon$ over all 100 repetitions of this process and assess the result versus $n$.  The results of this analysis are displayed in Figure \ref{fig:base_error}.

The error is fit with the exponential function
\begin{align}
    \label{eq:base_fit}
    \varepsilon(n) = c_1\left( 1 - e^{-\nicefrac{n^c_2}{c_3}}\right) \, .
\end{align}
Thus we obtain a baseline error on the order $10^{-4}$ when an excessively small time step is used.  This error \textit{only} occurs for extremely high sample rates, specifically those far beyond the necessary temporal resolution of the simulation.  For smaller sample rates, this error exponentially disappears.  For our simulations, we typically use a time step which would correspond to the low end of this scan, where the baseline error is on the order of $10^{-7}$ and can be safely ignored.  However,   this error is recovered in    the convergence relative to $dt$, as discussed in the next section.

\subsection{Convergence}

In order to examine the numerical validity of the model, we study its convergence with respect to several key ``artificial'' parameters: in particular, $n$ and $dt$.  We examine how well the simulation converges in the $n \to \infty$, $dt \to 0$ limits across a range of key physical parameters, in particular the linewidth $\Gamma$ of the frequency spectrum and the interaction strength (specified by $\alpha$ and the length scale $\xi$) in order to chart out the acceptable ranges for these numerical parameters (\ie  $n$ and $dt$) in the various physical regimes of the model.

Given the phenomenological behavior of the model for $\xi > 5$ is generally determined not by particular values of the interaction strength $\alpha$ or length scaling $\xi$, but by the overall interaction weight $\omega$, and in order to reduce computation time to a realistic level, we do not vary $\alpha$ and $\xi$ individually, but rather vary $\omega$ across a range of behavioral regimes and use a single pair of $\alpha$ and $\xi$ to represent the regime.
In studying the convergence with respect to $dt$ and $n$, we limit ourselves to isotropic interactions (i.e. $\alpha_x = \alpha_y$).  In doing so, we eliminate the error accrued from the approximations of the Hamiltonian using the Magnus expansion and can focus solely on the error that results from the ensemble size or time step.

\subsubsection{Convergence Relative to $n$}
\label{sec:n_conv}

There are several parameters which interact with the size of the ensemble in non-trivial ways which can affect the convergence of the echo behavior with respect to the number of particles $n$.  Given the interaction's dependence on a length scaling $\xi$, values of $n$ which are too small can artificially increase the interaction weight $\omega$ by cutting off the tails of the interaction.  In order to determine an appropriate value of $n$ such that the overall behavior of the echo is sufficiently convergent to the $n \to \infty$ limit, we compute echoes for a range of $\omega$ and $n$.

To test the convergence with respect to $n$, we use $\tau = \nicefrac{5}{2\Gamma}$ and vary $\omega$ from $1.5\, \Gamma$ up to $4.5\, \Gamma$.  This variation in $\omega$ typically leads to values of $\xi$ which have an upper bound of ten to fifteen lattice lengths.  To attempt to limit the potential for $n$ and $\xi$ to interact adversely while still scanning small ensembles, we use a minimum lattice size of $25 \times 25$ ($n = 625$).  We set the upper bound with a choice balanced between ensemble size and computational practicality, using a lattice of dimension $750$ and $n = 525,000$.  This upper bound of $n = 525,000$ will be used as the $n \to \infty$ approximation and the reference against which we calculate the error of the other echoes.

\begin{figure*}[!tbp]
    \centering
    \includegraphics[width=0.70\linewidth]{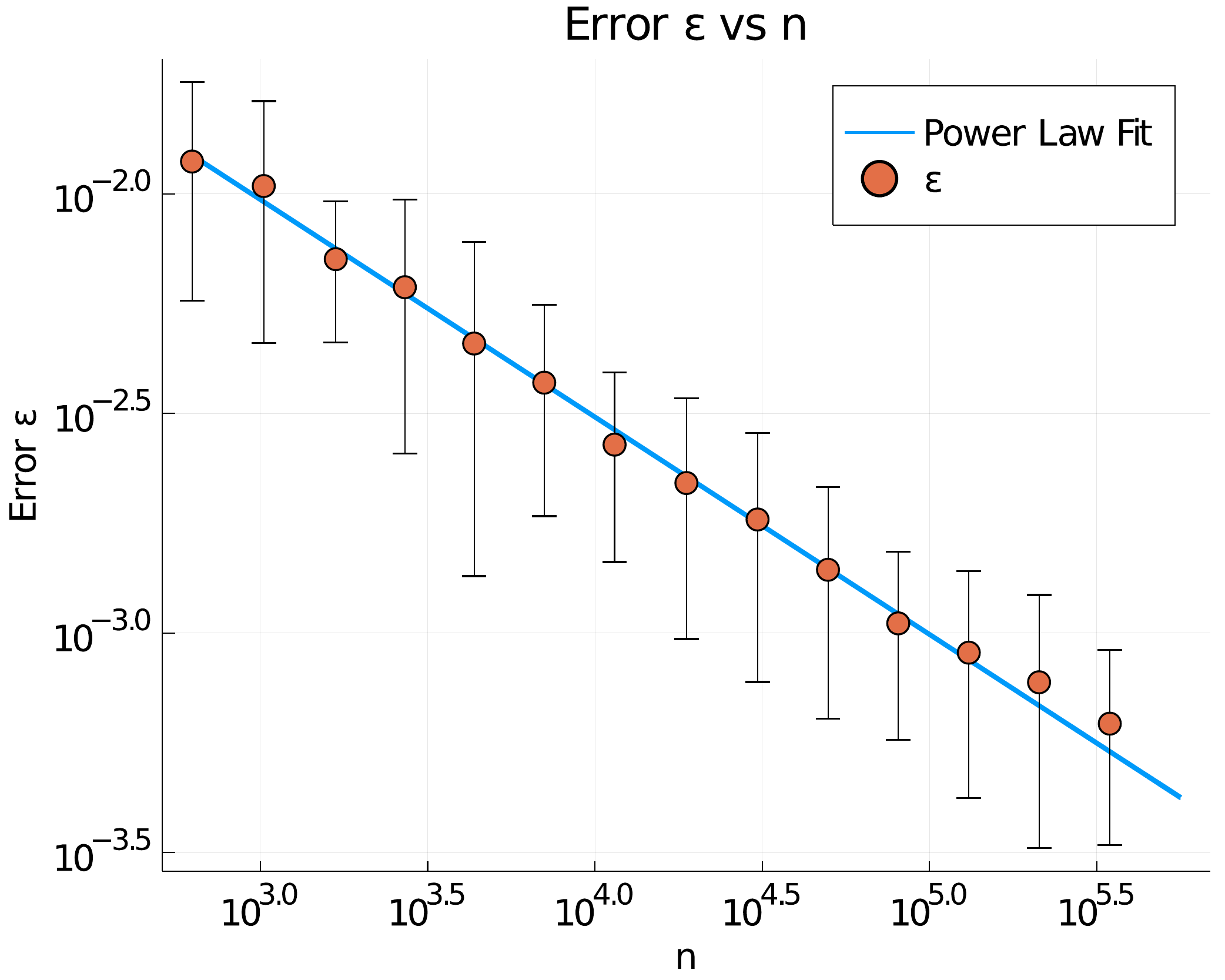}
    \caption{Average error magnitude $\varepsilon$ across all sampled $\omega$ vs $n$ with a power law fit.  Points represent the mean error across all $\omega$ at a given $n$ and error bars represent the standard deviation of the error for that ensemble size.}
    \label{fig:all_n_convergence}
\end{figure*}

A summary of the results of the convergence is presented in Figure \ref{fig:all_n_convergence}.
The average error is fit with a power law
\begin{align}
    \varepsilon = a n^{-b}\, .
\end{align}
Power law convergence is consistent with the discrete random sampling approach we apply to generate the frequencies of the ensemble.  The fit has a power $b = 0.495 \pm 0.035$, indicating that the model on average converges as $\nicefrac{1}{\sqrt{n}}$.

\begin{figure}[t]
    \centering
    \includegraphics[width=0.75\linewidth]{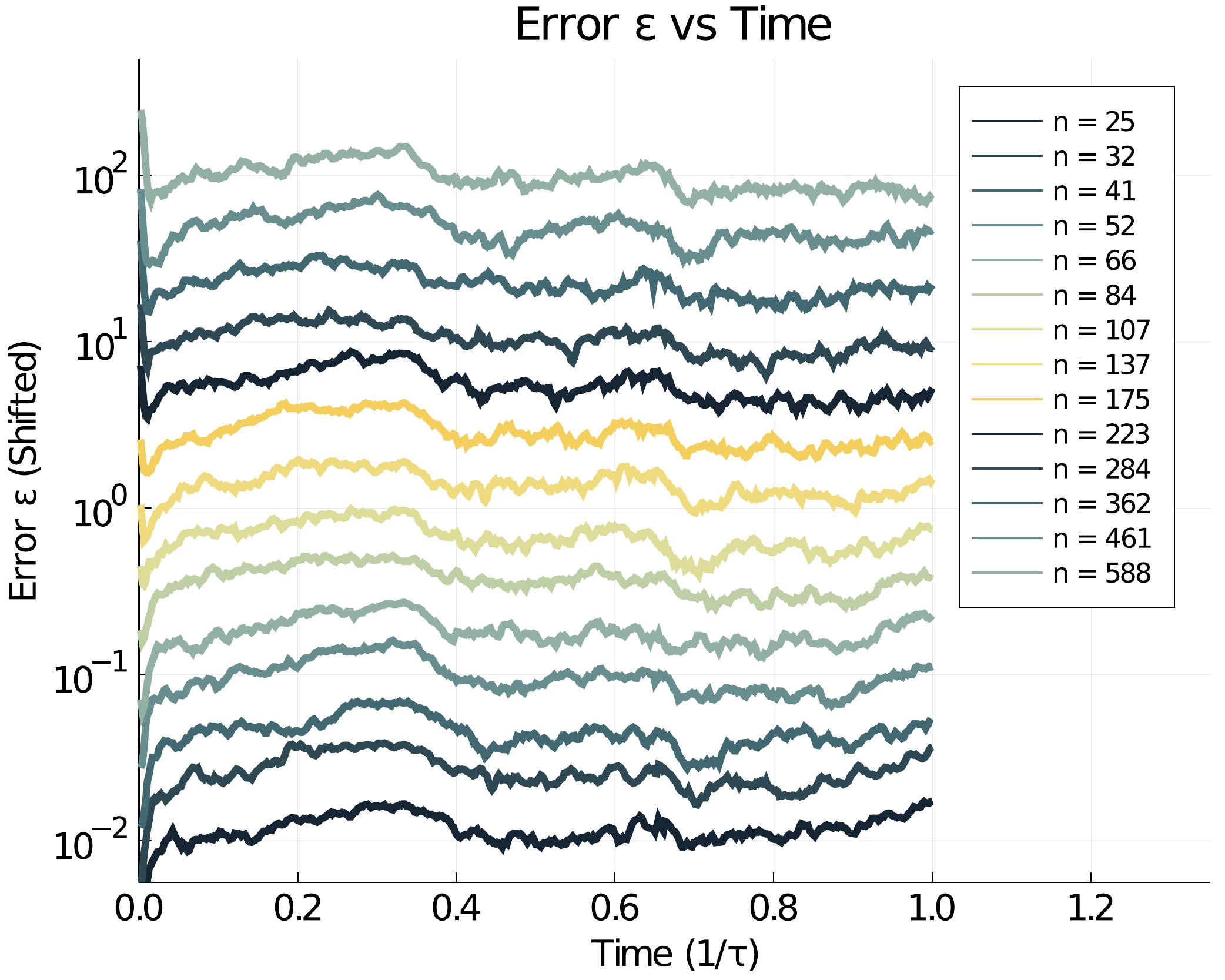}
    \caption{Error magnitude $\varepsilon$ (averaged over all sampled $\omega$) versus simulation time.  The error does not accumulate, i.e. the value at a time $t_1$ does not affect the error at a later time $t_2$.  To improve readability, error values have been scaled to create a linear offset on the log scale.}
    \label{fig:n_err_vs_t}
\end{figure}

Importantly, we note that the echo does not exhibit any ``drifting'', or accumulation of error, over the course of the echo.  The magnitude of the error near the end of the echo remains similar to the error near the beginning of the echo.  The error at a given point in time, averaged over all values of $\omega$, is given in Figure \ref{fig:n_err_vs_t}.

Based on these results, we use ensemble sizes of $n = 160,000$ when generating high quality echoes for direct analysis, or smaller ensembles such as $n = 10,000$ for generating large data sets where the broad phenomenology is the main goal (e.g. analysis of amplitudes) rather than direct analysis of the echoes and the lower computation time is necessary.  For a typical value of $\Omega dt \approx \nicefrac{1}{10}$, an ensemble of $n = 10,000$ results in computation times of around $4$ sec per echo and an ensemble of $n = 160,000$ of around $4$ minutes per echo.

\subsubsection{Convergence Relative to $dt$}
\label{sec:dt_conv}

The characteristic time scale of the dynamics of the model is set by the linewidth $\Gamma$ and the interaction strength $\omega$ (Equation \ref{eq:omega_def}), both of which contribute to the parameter $\Omega$, defined in Equation \ref{eq:big_omega_def}.  For our approximation of the time evolution to be accurate, $dt$ must take on values relative to $\Omega$ such that the condition $\Omega dt \ll 1$, as set in Equation \ref{eq:slowly_varying_condition}.  To appropriately chart out the accuracy of the model across the relevant parameters we compute echoes for a range of $\Omega$ and $dt$, and calculate their error relative to an echo with extremely small $dt$, which we use as an approximation of the $dt \to 0$ limit.

We sample time steps between $dt = {8\pi\times10^{-5}}/{\Gamma}$ and $dt = {\pi}/{5\Gamma}$. We sample $\omega$ from $1.5\, \Gamma$ up to $4.5\, \Gamma$, values which were empirically observed to cover the range of behaviors we are targeting.  Values of $\omega$ larger than $4.5 \,  \Gamma$ destroy the spin echo entirely.
We set the echo time $\tau$ as $\nicefrac{5}{2\Gamma}$.  This value is large enough for both a full decay of the FID in weakly interacting systems and for phase accumulation and notable echo perturbation in strongly interacting cases.  We utilize an ensemble of $n = 10,000$ spins for all trials.  To eliminate fluctuations in the error on account of the sampling, we employ a fixed frequency sampling for all echoes of a given linewidth.

The error for a given echo is calculated by considering the average magnitude of error across the echo when compared to the reference echo.  For each $\omega$ and $\Gamma$, we use the highest echo with the highest sampling rate (i.e. $dt = \nicefrac{4\times10^{-5}}{\Gamma}$) as an approximation of the $dt \to 0$ limit, and calculate the error for all other $dt$ with the same $\omega$ and $\Gamma$ relative to it.  Letting $S(t_i)$ denote the signal (absolute value) of the echo at time step $t_i$ and $S_{ref}(t_i)$ denote the same for the reference echo, the error $\varepsilon$ is given by
\begin{align}
    \label{eq:dt_err_def}
    \varepsilon = \frac{1}{L}\sum_{i = 1}^{L} \big| S(t_i) - S'_{ref}(t_i) \big| \, ,
\end{align}
for $L$ the number of points in the echo of interest.

In Equation \ref{eq:dt_err_def}, $S'_{ref}$ denotes the interpolated reference echo rather than the reference echo itself.  Given the difference in their sampling rates, $S_{ref}$ and $S$ have a different number of points.  In order to match the two echoes' sizes and calculate the error, we truncate $S_{ref}$ to have the same number of points as $S$, using linear interpolation between points of $S_{ref}$ to approximate its value at times that are between its recorded time steps.
This interpolation scheme introduces a potential source of artificial error, especially at points in the echo with high curvature.  However, the scale of this error is very small relative to the primary sources of error in the simulation, and it typically only becomes visible when comparing echoes both of which have very high sampling rates which are relatively similar in scale.

\begin{figure}[!tbp]
    \centering
    \includegraphics[width=0.69\linewidth]{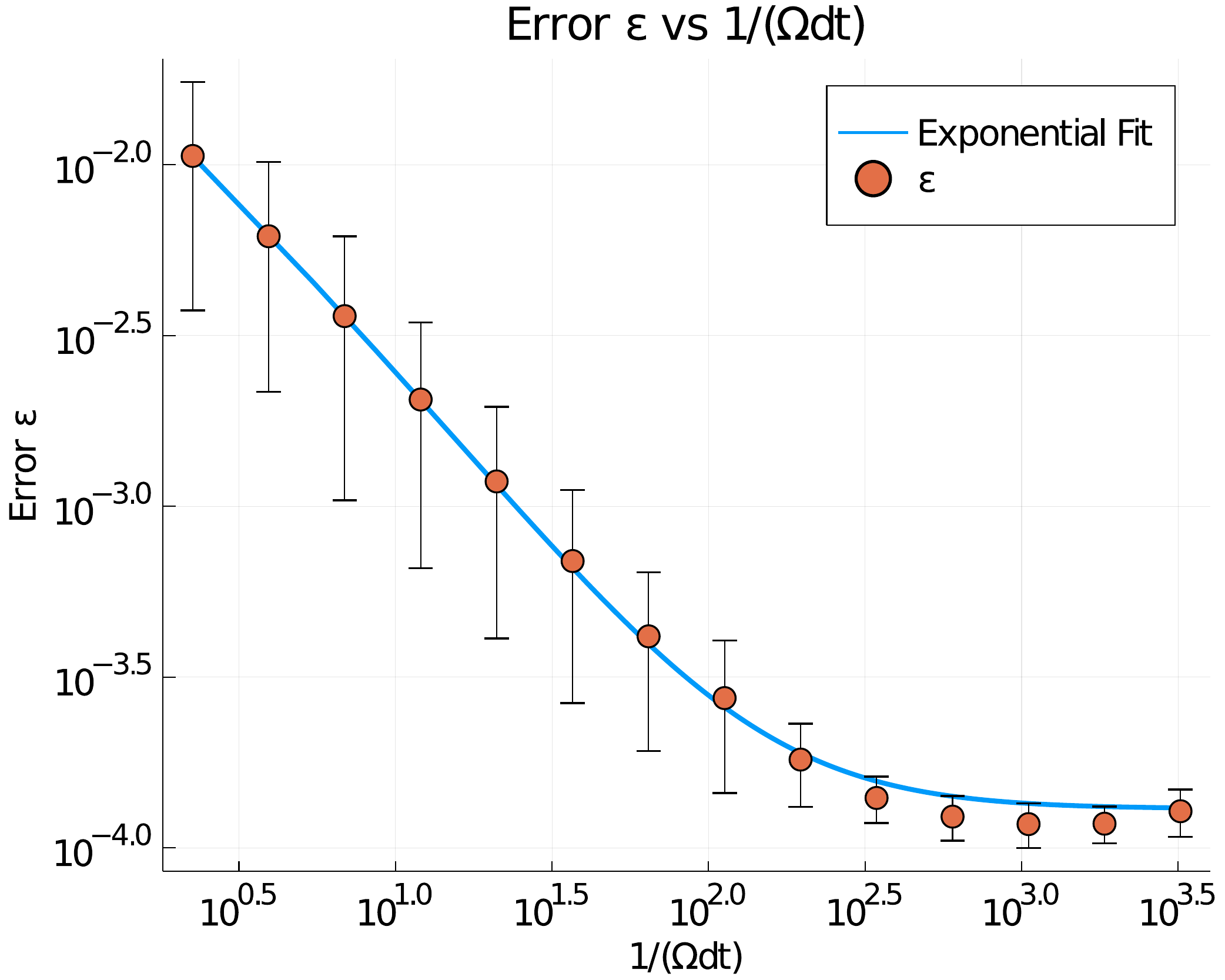}
    \caption{Average error magnitude $\varepsilon$ for all $\omega$ vs $\nicefrac{1}{\Omega dt}$.  Points represent the mean error across all $\omega$ at a given $\nicefrac{1}{\Omega dt}$ and error bars represent the standard deviation of the error for that sampling rate.}
    \label{fig:all_dt_convergence}
\end{figure}

The errors, averaged over all $\omega$, are presented in Figure~\ref{fig:all_dt_convergence}. 
For clarity, we group errors based off the approximate value $\Omega \approx \nicefrac{\Gamma}{\sqrt{2}}$, even though the sampled $\Omega$ vary by up to $15\%$ from this value.  The error bars denote the standard deviation in the values of $\varepsilon$ for a given $\nicefrac{1}{\Omega dt}$.

The mean value of $\varepsilon$ at each $\nicefrac{1}{\Omega dt}$ is fit using an exponential plus a constant:
\begin{align}
    \label{eq:dt_error_fit}
    \varepsilon = a e^{-\left(\nicefrac{x}{b}\right)^c} + d
\end{align}
for $x = \nicefrac{1}{\Omega dt}$.  The value of the baseline error $d$ is consistent with the numerical limits we observe in Section \ref{sec:baseline_error}.

\begin{figure*}[!tbp]

    \centering
    \includegraphics[width=0.73\linewidth]{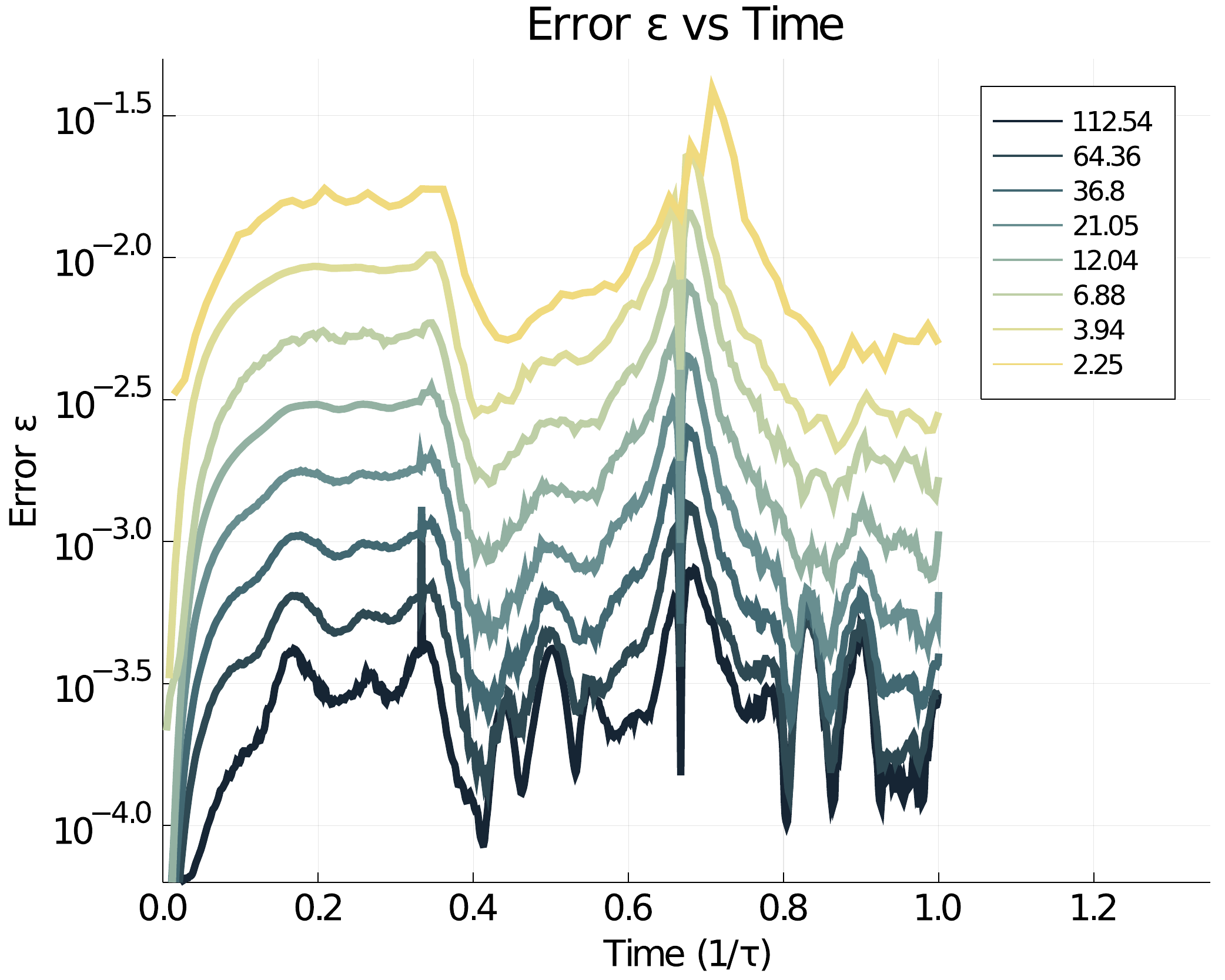}
    \caption{Error magnitude $\varepsilon$ (averaged over all $\omega$) versus time.  As in Section \ref{sec:n_conv}, the errors do not accumulate in time.  The legend indicates the value of $\nicefrac{1}{\Omega dt}$ for the given error curve.}
    \label{fig:dt_err_vs_t}
\end{figure*}

In the absence of $\omega$, the Hamiltonian is time independent and the approximations we use to time evolve the system become exact.
Therefore, as $\omega$ varies, the error caused by too large of a $dt$ should increase as well, which explains the overlap of the error bars between adjacent $\tfrac{1}{\Omega dt}$ samples in Figure~\ref{fig:all_dt_convergence}.
 Similarly to the case of $n$ convergence, there is no ``drift'' in the echo: the magnitude of the error does not tend to monotonically increase with time $t$ (see Figure \ref{fig:dt_err_vs_t}).

In summary, the model shows exponential convergence with respect to the choice of time step $dt$. 
The accuracy of the simulation relative to the choice of $dt$ depends only on its value relative to the interaction $\omega$.
In order to guarantee an error of less than $0.1\%$, depending on the value of $\omega$, \ie   $\nicefrac{1}{\Omega dt}$ must range between $5$ and $20$.  We used values of $\nicefrac{1}{\Omega dt} \approx 10-15$, which results in a computation time of 4 sec per echo for $n = 10,000$.

\section{CONCLUSIONS}
\label{sec:conclusion}

Having explored the dynamics of the interaction across a variety of its spatial  ranges $\xi$ and strengths $\omega$, we can begin to see the applications of this model to a real system.  In order to observe any significant asymmetry in the echo, an interaction strength $\omega > 2.5\Gamma$ is typically required.  The strong multi-peak behavior only begins to emerge when $\omega > 3\, \Gamma$.  Observing this behavior in a material with small $\xi$ ($\approx 2$) is unlikely, as the presence of $\alpha_z$ can cause strong dissipation and our mean field approximation is less accurate at small $\xi$.

In the intermediate range of $\xi$, the echo and spectral behavior differ from the global limit but the mean field approximation is likely accurate.
Nearest- and next nearest-neighbor terms do not dominate, and the dissipative effects are mitigated at the interaction strengths necessary to induce the notable pulse-dependent shifts.
Additionally, the required interaction strength is significantly reduced on account of the larger number of spins which contribute to the mean field magnetization.
For $\xi = 6$, only a coupling of $\alpha > \Gamma/50$ is necessary.  As $\xi$ approaches the long range regime $\xi \to 20$, the coupling needs only be on the order of $\Gamma/1000$.
In a scenario where the behavior described here is experimentally observed, we expect it is driven by long range correlations, where $\xi > 10.0$ and $\alpha$ need only be a fraction of the linewidth, anywhere from $1\%$ to $0.1\%$ of $\Gamma$ depending on the range of the interaction.

The most valuable tool in identifying the interaction's strength is the pulse dependent shift~\cite{theory_paper}.  By adjusting the orientation of the crystal axes of the sample relative to the constant magnetic field $B_0$ and measuring the magnitude of the pulse dependent shift, one can extract the differences between different pairings of the interaction strengths.  The ability to measure the relative strengths of the interaction along different spin-axes represents a potent probe for the anisotropy of the electronic spin susceptibility.  By combining this information with measurements of the anisotropy of the hyperfine interactions, as outlined for example in  \mbox{Refs. \cite{Vachon08,Curro09}}, one can extract previously unavailable information about the electronic structure and give valuable insight into the structure of strongly correlated systems.

\section{CRediT author statement}

\noindent{\bf Charles Snider}: Investigation, Software Programming - CUDA and Julia, Development and Validation, Formal analysis, Visualization, Data curation, Writing-Original draft preparation. 
{\bf Stephen Carr}: Investigation, Software Programming, Development and Validation, Testing, Formal analysis, Visualization, Writing-Original draft, Writing-Reviewing and Editing. 
{\bf Dmitri E. Feldman}: Methodology, Investigation, Supervision, Formal analysis, Resources, Writing-Reviewing and Editing. 
{\bf  Chandrasekhar Ramanathan}: Methodology, Development and Validation,  Reviewing and Editing. 
{\bf Vesna F. Mitrovi{\'c}}: Conceptualization, Methodology, Supervision, Formal analysis, Visualization, Resources, Writing-Reviewing and Editing.

\section*{Acknowledgments}
We thank Prof. Brad Marston for helpful helpful advice during  the development of the program. 
This work was supported by the National Science Foundation under grant No. OIA-1921199. The calculations were conducted using computational
resources and services at the Center for Computation and Visualization, Brown University.

\bibliographystyle{elsarticle-num}
\bibliography{refs.bib}

\end{document}